\documentclass[aps,pra,twocolumn,letterpaper,showpacs,superscriptaddress,floatfix,10pt]{revtex4-1}
\usepackage[pdftex]{graphicx}
\graphicspath{ {./images/} }
\usepackage{dcolumn}
\usepackage{bm}
\usepackage{bbm}
\usepackage[usenames, dvipsnames]{color}
\usepackage{comment}
\usepackage[colorlinks, linkcolor=blue]{hyperref}

\usepackage{layouts}

\usepackage[T1]{fontenc}
\usepackage{exscale}
\usepackage{amsbsy}
\usepackage{subfigure}
\usepackage{textcomp}
\usepackage{physics}
\usepackage{mathtools}
\usepackage{amsfonts}
\usepackage{amssymb}
\usepackage{amsmath}
\usepackage{mathtools}
\usepackage{tensor}

\usepackage{amsthm}
\theoremstyle{remark}

\usepackage{multirow}

\usepackage[scr=boondoxo, scrscaled=1.05]{mathalfa}

\usepackage{subfigure}
\usepackage{overpic}

\usepackage{array}
\newcolumntype{L}[1]{>{\raggedright\let\newline\\\arraybackslash\hspace{0pt}}m{#1}}
\newcolumntype{C}[1]{>{\centering\let\newline\\\arraybackslash\hspace{0pt}}m{#1}}
\newcolumntype{R}[1]{>{\raggedleft\let\newline\\\arraybackslash\hspace{0pt}}m{#1}}
\usepackage{tabularx}

\def\KeyWord#1{$\backslash$\IfColor{$\!\!$\textRed{#1}\textBlack}{#1}$\!\!$}

\usepackage{wasysym}

\def\ket#1{|#1\rangle}

\begin{document}
\title{Absence of disordered Thouless pumps at finite frequency}

\author{Dominik Vuina}
\email{dominikv@bu.edu}
\affiliation{Department of Physics, Boston University, Boston, Massachusetts 02215, USA}
\author{David M. Long}
\affiliation{Department of Physics, Boston University, Boston, Massachusetts 02215, USA}
\affiliation{Condensed Matter Theory Center and Joint Quantum Institute, Department of Physics, University of Maryland, College Park, Maryland 20742, USA}
\author{Philip J. D. Crowley} 
\affiliation{Department of Physics, Harvard University, Cambridge, Massachusetts 02138, USA}
\author{Anushya Chandran}
\affiliation{Department of Physics, Boston University, Boston, Massachusetts 02215, USA}

\date{\today}

\begin{abstract}    
    A Thouless pump is a slowly driven one-dimensional band insulator which pumps charge at a quantised rate.
    Previous work showed that pumping persists in weakly disordered chains, and separately in clean chains at finite drive frequency. 
    We study the interplay of disorder and finite frequency, and show that the pump rate always decays to zero due to non-adiabatic transitions between the instantaneous eigenstates.
    However, the decay is slow, occurring on a time-scale that is exponentially large in the period of the drive.
    In the adiabatic limit, the band gap in the instantaneous spectrum closes at a critical disorder strength above which pumping ceases.
    We predict the scaling of the pump rate around this transition from a model of scattering between rare states near the band edges.  
    Our predictions can be experimentally tested in ultracold atomic and photonic platforms.
     
\end{abstract}

\maketitle

\emph{Introduction.}---Periodically driving a one-dimensional band insulator can lead to Thouless charge pumping---transporting charge at a quantised rate across the system in the adiabatic limit~\cite{PhysRevB.27.6083_tholuess_main,citro_aidel_nature_rev}. Recent advances in experimental realization of Thouless pumps in cold atom~\cite{thouless_pump_cold_atoms_exp_fermions,thouless_pump_cold_atoms_exp_bosons, PhysRevLett.116.200402_exp_cold_atoms,disorder_thouless_pump_cold_atoms_exp,PhysRevLett.129.053201_ultracold_atoms,interactions_exp_cold_atoms} and photonic~\cite{PhysRevLett.109.106402_photo_exp, PhysRevB.91.064201_photo_exp,Ke_2016_exp_photonics,PhysRevLett.117.213603_interactions_exp_photonics,dissipation_exp_photonics1,disordered_thouless_pump_photonics_exp, interactions_exp_photonics, dissipation_exp_photonics2} systems have prompted studies of the robustness of this dynamical phenomenon to disorder~\cite{niu1984quantised, PhysRevB.35.2188_exponentially_small_corr,thouless1998topological_exponentially_small_corr,QIN20162317_disorder,PhysRevB.100.184304_disorder,PhysRevLett.123.266601_disorder, disorder_thouless_pump_cold_atoms_exp, disordered_thouless_pump_photonics_exp, PhysRevLett.124.086602_disorder,PhysRevResearch.2.042035_disorder,PhysRevA.101.052323_disorder, PhysRevA.103.043310_disorder, PhysRevA.106.L051301_disorder,grabarits2023floquetanderson}, interactions~\cite{PhysRevB.68.165312_interactions,PhysRevLett.106.110405_interactions,PhysRevA.84.013608_interactions,PhysRevLett.109.257201_interactions,PhysRevA.90.053623_interactions,PhysRevB.94.235139_interactions,PhysRevX.7.011018_interactions,PhysRevB.98.245148_interactions,PhysRevA.99.053614_interactions,PhysRevA.101.053630_interactions,interactions_exp_cold_atoms,niu1984quantised, interactions_exp_photonics,PhysRevResearch.2.042024_interactions,esin2022universal_interactions,PhysRevB.106.045141_interactions,PhysRevLett.117.213603_interactions_exp_photonics}, finite frequency driving~\cite{PhysRevLett.64.1812_non_adiabatic,PhysRevB.50.11902_non_adiabatic,avron1999quantum_non_adiabatic, PhysRevLett.120.106601_non_adiabatic, 10.21468/SciPostPhys.12.6.203_non_adiabatic} and dissipation~\cite{dissipation_exp_photonics1, dissipation_exp_photonics2, arceci2020dissipation}. Previous works have considered non-adiabatic effects in the clean Thouless pump~\cite{PhysRevLett.120.106601_non_adiabatic} and the effects of disorder in the adiabatic limit~\cite{PhysRevLett.123.266601_disorder,grabarits2023floquetanderson}. 

\begin{figure}
    \centering
    \includegraphics[width=\linewidth]{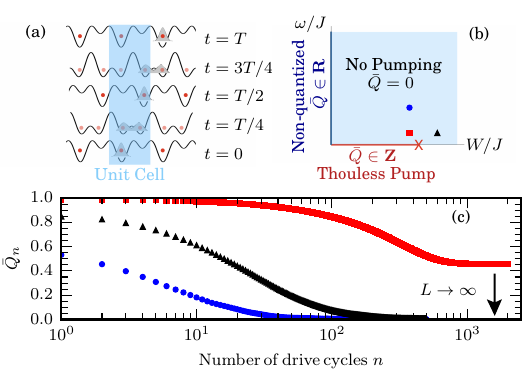}
    \caption{(a) Cartoon of the Thouless pump---charge is transported through one unit cell per period by periodically varying the potential. (b) Phase diagram of the disordered Thouless pump with bounded disorder. Quantised pumping occurs in the adiabatic limit $\omega \to 0$ for weak disorder $W<W_c$. At finite frequency and disorder, the pump rate $\bar{Q}$ is zero. In the clean limit $W=0$ the pump rate assumes a finite, non-quantised value. (c) The charge pumped per cycle in model~\eqref{eq:model_eq}, averaged over the first $n$ drive cycles, $\bar{Q}_n$ decays to zero in the thermodynamic limit for any $\omega, W > 0$. Saturation values of $\bar{Q}_n$ give the pump rate $\bar{Q}$ in Eq.~\eqref{eq:charge_pump_per_period}.  \emph{Parameters} of the model~\eqref{eq:model_eq} are $J=1.0$, $\Delta_0=1.5J$, $\delta_0=0.5J$ and $L=350$, and $( W/J, \omega/J) = (3.0, 0.3)$ (red), $(W/J, \omega/J) = (3.0, 1.0)$ (blue) and $(W/J, \omega/J) = ( 5.5, 1.0)$ (black). The $(W/J, \omega/J)$ values are marked in (b).}
    \label{fig:phase_diag}
\end{figure}

This Letter shows that the interplay of disorder and finite frequency driving causes the charge pumping rate to vanish at long times. We uncover the two dimensional phase diagram of the Thouless pump as a function of the frequency of the drive $\omega$ and disorder strength $W$ (Fig.~\ref{fig:phase_diag}(b)). The phases are distinguished by the behaviour of the \emph{pump rate} $\bar{Q}$---the average charge pumped per period at long times.

As first shown by Thouless~\cite{PhysRevB.27.6083_tholuess_main}, the clean $(W=0)$ model in the adiabatic limit $(\omega \to 0)$ exhibits quantised charge pumping $\bar{Q} \in \mathbb{Z}$ when initialised in its instantaneous ground-space at integer filling. At low, but finite frequency, non-adiabatic effects cause an $\mathcal{O}(\omega^2)$ decrease in the pump rate, which is consequently no longer quantised,  $\bar{Q}\in \mathbb{R}$~\cite{PhysRevLett.120.106601_non_adiabatic}.

At weak disorder charge pumping is robust only in the adiabatic limit. For models with bounded disorder, the charge pumping is quantised $\bar{Q} \in \mathbb{Z}$ below a critical disorder strength $(W<W_c)$~\cite{niu1984quantised}. As the disorder strength is increased above its critical value the pump rate drops discontinuously to zero, $\bar{Q}=0$. 

For any finite drive frequency and disorder strength $(\omega, W > 0)$, we show that the system exhibits only transient charge pumping. The pumping decays on a finite time scale $\tau_Q$ such that $\bar{Q}=0$ (Fig.~\ref{fig:phase_diag}(c)).

We identify two mechanisms for pumping decay, and their associated time scales $\tau_Q$, which apply in the adiabatic and non-adiabatic cases. 
In the adiabatic limit, below the critical disorder strength, the instantaneous eigenstates split into two bands of Anderson localised states~\cite{PhysRev.109.1492_anderson_loc, lee1985disordered} which pump in opposite directions. Any scattering between these bands causes the pumping to decay to zero. 

At finite frequency there are non-adiabatic transitions between the  bands. A Landau-Zener analysis predicts a pumping decay time which is exponentially long in the period of the drive $\tau_Q = \mathcal{O}(e^{\alpha/\omega})/\alpha$, where $\alpha$ is an energy scale related to the band gap (Fig.~\ref{fig:phase_diag}(c)). Relating this timescale to a length scale allows for verification of our prediction for $\tau_Q$ through a finite size scaling analysis of $\bar{Q}$ as $L\to \infty$, $\omega \to 0$.

In the adiabatic limit the instantaneous band gap closes at least once during a period for $W\geq W_c$. The states of the two bands form avoided level crossings with gaps scaling with the system size. This allows transitions between the bands at any drive frequency. The associated timescale is related to the density of states at the edges of the bands, $\tau_Q = \mathcal{O}(e^{R/\sqrt{W-W_c}})/\omega$. Similarly to the non-adiabatic mechanism, we verify this prediction with a finite size scaling analysis of $\bar{Q}$ near the $W=W_c$ transition point in the adiabatic limit. 

\emph{Model}.---We work with the disordered Rice-Mele model~\cite{PhysRevLett.49.1455_rm_model}. This is a time periodic model $(H(t+T) = H(t))$ of spinless fermions on a $N=2L$ site chain, of $L$ unit cells, and is given by
\begin{equation}\label{eq:model_eq}
    \begin{split}
        H(t) = - \sum_{j=0}^{L-1} \left( J_+(t) c^{\dagger}_{2j-1} c_{2j} + J_-(t) c^{\dagger}_{2j} c_{2j+1} + \mathrm{h.c.} \right) \\
     - \sum_{l=0}^{N-1} \left( (-1)^l\Delta(t) + W \zeta_l \right) c^{\dagger}_{l} c_{l},
    \end{split}
\end{equation}
where $c^{\dagger}_j$ creates a fermion on site $j$. Parameters $J_{\pm}(t)=J \pm \delta_0 \cos{\omega t}$ represent intra/inter unit cell hopping amplitudes and $\Delta(t)=\Delta_0 \sin{\omega t}$ is the on site potential. $W\zeta_l$ represents the on site disorder of strength $W$, with independently and identically distributed $\zeta_l$ drawn uniformly from $[-1/2, 1/2]$. We restrict to periodic boundary conditions. 

The pump rate $\bar{Q}$ is given by
\begin{equation}\label{eq:charge_pump_per_period}
    \bar{Q}=\lim_{n \to \infty} \frac{1}{n}\int_0^{nT} \, \mathrm{d}t \, \langle \psi(t)|I(t)| \psi(t) \rangle,
\end{equation}
where 
\begin{equation}\label{eq:curr_density}
    I(t)=\frac{i}{L}\sum_{j=0}^{L-1} \left( J_+(t)c^{\dagger}_{2j-1} c_{2j} + J_-(t) c^{\dagger}_{2j} c_{2j+1} - h.c.\right)
\end{equation}
is the current operator. Here $\ket{\psi(t)}$ is the time evolved many-body ground state of $H(t=0)$ at half-filling. 

We use the Floquet diagonal ensemble to calculate the pump rate
\begin{equation}\label{eq:Q_floquet_diag}
    \bar{Q} = \sum_{m=0}^{N-1} p_m \int_0^T \, \mathrm{d}t \, I_{mm}(t),
\end{equation}
where $m$ enumerates the Floquet states, defined by $U^{\dagger}(t,0)f_m(t)U(t,0) = f_m(t) e^{-i \epsilon_m t}$ with $f_m(t)=\sum_j a_{m_j}(t)c_j$ annihilating the $m^{\mathrm{th}}$ Floquet state and $f_m(t+T)=f_m(t)$. The Floquet state populations are $p_m=\langle \psi(0)|f_m^{\dagger}(0)f_m(0)|\psi(0)\rangle$ and $I_{mn}(t)$ are the corresponding matrix elements of $I(t) = \sum_{mn} I_{mn}(t)f_m^{\dagger}(t)f_n(t)$. Here $|\psi(0) \rangle$ denotes the ground state of $H(t=0)$ at half-filling.  

We now describe the properties of the single particle eigenstates of the Rice-Mele model. The instantaneous spectrum of the model~\eqref{eq:model_eq} in the clean $W=0$ limit splits into two bands of delocalised states separated by a gap. With weak random disorder the instantaneous eigenstates remain organised in two bands; however, they are now Anderson localised~\cite{PhysRev.109.1492_anderson_loc, lee1985disordered}. We assume that disorder broadens the clean system bands such that the four edges of the bands become~\cite{ edwards1971regularity_anderson_den_states,lacroix1981density_of_state_anderson}
\begin{equation}\label{eq:band_edge}
 \begin{split}
     E_b=\left(E_{+, \uparrow}+\frac{W}{2}, E_{+, \downarrow}-\frac{W}{2}, E_{-, \uparrow}+\frac{W}{2}, E_{-, \downarrow}-\frac{W}{2}\right),
 \end{split}
\end{equation} 
where $E_{+(-), \uparrow (\downarrow)}$ refers to the top and bottom clean band edges of the upper and lower bands. With random disorder the edges of the bands have Lifshitz tails~\cite{biroli2010anderson_lifshitz_tails,PhysRevLett.109.076402_lifshitz_tails, PhysRevB.93.075120_lifshitz_tails}. The states in these tails are delocalised over clusters of contiguous sites with low disorder (close to extremal values of onsite potential $\zeta_l=\pm 1/2$). In one dimension Lifshitz tails have the density of states~\citep{doi:10.1080/00018736400101061,lifshitz1965energy,kramer1993localization, PhysRevLett.109.076402_lifshitz_tails}
\begin{equation}\label{eq:den_states_tails}
 n(E) \propto \exp\left(-R|E_b-E|^{-1/2}\right),
\end{equation}
where $E_b$ are the edges of the bands~\eqref{eq:band_edge} and $R$ is some scale with units of $\sqrt{E}$.

In the adiabatic limit of the model~\eqref{eq:model_eq}, the quantised charge pumping is intuitively explained using the instantaneous basis. The system is initialised in its many-body instantaneous ground state at half filling---an occupied lower band. Adiabatic following of this instantaneous eigenstate in a period of the drive results in quantised pumping of charge along the chain. This occurs if there is a nontrivial spectral flow of instantaneous single particle eigenstates, which may be related to a nonzero Chern number $C$ in a 2D synthetic lattice (Appendix C). As the model~\eqref{eq:model_eq} is local, the spectral flow of the two bands must cancel. The Chern numbers of the instantaneous bands in model~\eqref{eq:model_eq} is $C=\pm 1$, with current flowing in the opposite direction along the chain if the upper band is initially occupied.

\begin{figure}
    \centering
    \includegraphics[width=\linewidth]{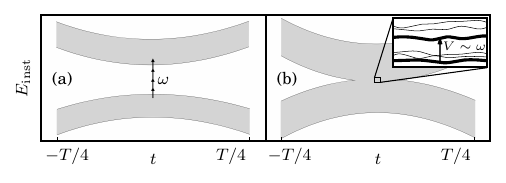}
    \caption{Cartoons of pumping decay mechanisms. (a) Non-adiabatic scattering mechanism. At $0<W<W_c$ a non-zero frequency gives a finite transition probability per period between the states belonging to the two bands---modelled as a parabolic level crossing. Shaded regions represent bands of instantaneous eigenstates. (b) Adiabatic scattering mechanism. The instantaneous gap closes at least once within the period (here at $t=0$). The gaps between the states in the overlap region are set by their matrix elements $V$. When $V=\mathcal{O}(\omega)$ the probability of scattering to the opposite band is $\mathcal{O}(1)$.}
    \label{fig:mechanisms}
\end{figure}

The breakdown of quantised pumping away from the adiabatic limit, or at strong enough disorder $W \geq W_c$, is understood through processes connecting the bands. Any finite rate of scattering between the bands causes a decay in quantised pumping. The pump rate vanishes, $\bar{Q}=0$, when the population of the two bands becomes equal. This can occur either by non-adiabatic scattering between the instantaneous eigenstates belonging to the two bands at finite drive frequencies $\omega > 0$, or via the instantaneous spectral gap closing in the adiabatic limit at $W=W_c$, the critical disorder strength, which is given by $E_b=0$ in Eq.~\eqref{eq:band_edge}. 

\emph{Non-adiabatic transition mechanism.}---At weak disorder $W<W_c$ and finite drive frequency, the non-adiabatic transitions between the bands (Fig.~\ref{fig:mechanisms}(a)) destroy pumping. The instantaneous band gap does not close within a period of the drive. A Landau-Zener analysis predicts a transition probability between any two instantaneous eigenstates, which is exponentially small in the drive frequency $\mathcal{O}\left(e^{-\alpha/\omega} \right)$. 
Adding up the contributions from the transitions between all the states belonging to the two bands gives a decay time scale for $\bar{Q}$ with an initially populated lower band. We relate this timescale to the scaling form of $\bar{Q}(L)$ at finite system sizes, allowing for finite size scaling analysis.

In detail, we model the transitions between any two instantaneous eigenstates in the upper and lower bands respectively during a period with a parabolic level crossing model~\cite{SUOMINEN1992126_lz, PhysRevA.86.033415_lz,kam2020analytical_lz}. At times when the instantaneous spectral gap is smallest ($t^*=2 mT$, for $m \in \mathbb{N}$) the instantaneous energies of Eq.~\eqref{eq:model_eq} are of the form $E_{\pm}(t) \approx \pm \left( \epsilon_0 + \kappa \omega^2(t-t^*)^2\right)$ , where $\epsilon_0$ is the energy scale of the separation between the states belonging to opposite bands and $\kappa$ quantifies their curvature at the minimum (Fig.~\ref{fig:mechanisms}, Appendix B). The matrix element connecting the localised instantaneous eigenstates is $V \sim J\exp \left( -|r_+ - r_-|/\max\{\xi_{\mathrm{loc}}(E_+), \xi_{\mathrm{loc}}(E_-)\} \right)$, where $r_{\pm}$ denote localisation centers of the instantaneous eigenstates in the upper and lower bands respectively, and $\xi_{\mathrm{loc}}(E_{\pm})$ are their localisation lengths. The scattering event is treated within a two level model

\begin{equation}\label{eq:effective_two_state_problem}
\begin{split}
    H^{\mathrm{eff}}(t-t^*) = \left(\epsilon_0 + \kappa\omega^2 (t-t^*)^2\right)\sigma_z + V\sigma_x + \\ \mathcal{O}\left( \omega^3 (t-t^*)^3\right).
\end{split}
\end{equation}
Adding up the contributions from all the states in the two bands gives the excitation rate between the bands
\begin{equation}\label{eq:exc_rate}
\begin{split}
    \Gamma =\omega\int \mathrm{d}r_+ \, \mathrm{d}r_- \, \mathrm{d}E_+ \, \mathrm{d}E_- \, n(E_+) \, n(E_-)  \\
    \times P_{\mathrm{exc}}(\epsilon_0=|E_+-E_-|, \omega, V),
\end{split}
\end{equation}
where $E_{\pm}(t^*)$ are the instantaneous energies of states at time $t=t^*$ in the respective bands and $n\left( E_{\pm} \right)$ are their densities of states. The probability of excitation $P_{\mathrm{exc}}$ is calculated from the parabolic crossing model~\eqref{eq:effective_two_state_problem}.

In the limit of small drive frequencies $\epsilon_0, V, \kappa \gg \omega$ and large instantaneous energy gap $\epsilon_0 \gg V$, the dominant contribution to the excitation rate~\eqref{eq:exc_rate}
comes from the Lifshitz tail states near the band edges (Fig.~\ref{fig:mechanisms}(a)). The reason is as follows.  Eq.~\eqref{eq:band_edge} implies that the smallest instantaneous band gap $\epsilon_0 \geq W_c - W$. Meanwhile, the matrix elements are exponentially small in the distance between the localisation centers of the two states. Therefore, $\epsilon_0 \gg V$ holds typically for $W$ not too close to $W_c$. In this limit the transition probability from Eq.~\eqref{eq:effective_two_state_problem} is~\cite{SUOMINEN1992126_lz} 
\begin{equation}\label{eq:trans_prob_non_ad}
    P_{\mathrm{exc}}(\epsilon_0, \omega, V) = \frac{\pi V^2}{2 \omega \sqrt{\kappa \epsilon_0}} \exp(-\frac{8}{3}\frac{\epsilon_0^{3/2}}{\omega \sqrt{\kappa}}).
\end{equation}
In the small drive frequency limit $\epsilon_0 \gg \omega$, only the states near the band edges---with the smallest gap $\epsilon_0$---contribute significantly to the scattering rate~\eqref{eq:exc_rate}. Moreover, near the band edges the localisation length smoothly depends on the energy, so we approximate $\xi_{\mathrm{loc}}(E_{\pm})=\xi_{\mathrm{loc}}$.

Integrating over the distances and energies in the two bands, with the density of states in Eq.~\eqref{eq:den_states_tails} and transition probability~\eqref{eq:trans_prob_non_ad} we obtain the excitation rate (Appendix A)
\begin{equation}
    \Gamma \propto L \frac{J^2}{\sqrt{\kappa(W-W_c)}} \exp\left(- k J/\omega - \mathcal{O}\left(\omega^{-1/3}\right)\right),
\end{equation}
where $L$ is the number of unit cells. The total excitation rate is extensive since the time-dependent part of the Hamiltonian is also extensive. The $\omega^{-1/3}$ contribution in the exponent, coming from the competing density of states and transition probability near the band edges, is negligible at small drive frequencies.

The excitations between the two bands in subsequent cycles are incoherent in the thermodynamic limit. To see this note that, with random disorder, the charges experience a different uncorrelated potential after being shifted by one lattice site every period on average. Dynamical phases accrued between the subsequent transitions are thus also uncorrelated---allowing for incoherent treatment of these excitations. Initialising the system in the instantaneous lower band, the inverse of the excitation rate per state sets the timescale of decay of quantised charge pumping
\begin{equation}\label{eq:omega_loc_time}
     \tau_Q \propto \frac{\sqrt{\kappa(W-W_c)}}{J^2}e^{kJ/\omega}. 
\end{equation}
A result of this long decay timescale is that if the disordered pump is observed on time scales $t \ll \tau_Q$ it appears stable to non-adiabatic effects. Therefore, quantised pumping in the adiabatic $(\omega \to 0)$ limit only occurs when $t/\tau_Q \to 0$ as $t, \tau_Q \to \infty$. 

In finite sized systems, the decay is cut off by coherence effects in the pump dynamics. As the drive protocol is periodic, charges encounter the same disorder potential after moving $L$ unit cells around the chain. The dynamical phases accumulated between excitations in subsequent cycles become correlated on timescales $t>LT$. These coherence effects in the population dynamics of the two bands result in their unequal occupation at long times. Since the two bands carry opposite currents, the pump rate is finite $|\bar{Q}|>0$.

\begin{figure}
    \centering
    \includegraphics[width=\linewidth]{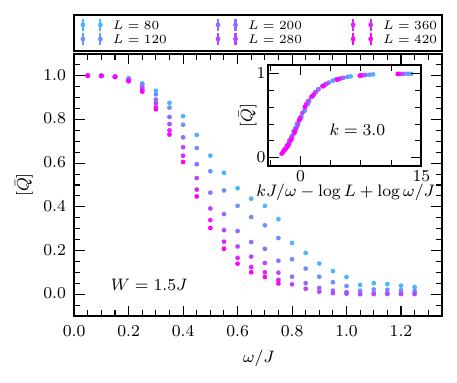}
    \caption{Disorder averaged $\bar{Q}$ as a function of drive frequency at different system sizes. The region of quantised pumping $[\bar{Q}] = 1$ shrinks as $L \to \infty$. Inset shows the scaling collapse in Eq.~\eqref{eq:scaling_form_omega} with $\xi_{\omega}$ in Eq.~\eqref{eq:omega_loc_length}. \emph{Parameters}: $J=1.0$, $\Delta_0=1.5J$, $\delta_0=0.5J$, $W=1.5J$, $k=3.0$ and $[\bar{Q}]$ is averaged over 200 disorder realisations.}
    \label{fig:scaling_fig}
\end{figure}

Finite size scaling analysis of the pump rate $\bar{Q}$, can be used to confirm Eq.~\eqref{eq:omega_loc_time}. It is natural to compare the system size $L$ to the length scale over which charge is pumped across the system $\xi_{\omega}$. Since the charges, on average, move through one unit cell per period, charge is transferred across
\begin{equation}\label{eq:omega_loc_length}
    \xi_{\omega} = \frac{\tau_Q}{T} \sim \frac{\omega}{J} \frac{\sqrt{\kappa(W-W_c)}}{J}e^{kJ/\omega}
\end{equation}
unit cells. Figure~\ref{fig:scaling_fig} indeed shows the disorder averaged $[\bar{Q}]$ is consistent with the scaling form 
\begin{equation}\label{eq:scaling_form_omega}
    [\bar{Q}](\xi_{\omega}, L) \sim \Tilde{Q}(\xi_{\omega}/L).
\end{equation}
In the small drive frequency limit, $\xi_{\omega}$ is dominated by the exponential dependence on $\omega$---plotting $\log(\xi_{\omega}/L)$ on the horizontal axis gives a good data collapse, as shown in the inset of Fig.~\ref{fig:scaling_fig}. The dimensionless parameter $k$ in the exponent of Eq.~\eqref{eq:omega_loc_length} is left as a free parameter, with the leading order prediction as in Appendix A.

\emph{Adiabatic transition mechanism.}---The adiabatic scattering mechanism (Fig.~\ref{fig:mechanisms}(b)) controls the transition on the $\omega \to 0$ line. When $W>W_c$, the two instantaneous bands overlap in an energy window $W-W_c$ for part of the period (Fig.~\ref{fig:mechanisms} (b)). The pertinent instantaneous eigenstates are, again, the Lifshitz states in the overlapping edges of the bands. 

At any drive frequency the probability of transition between these states is $\mathcal{O}(1)$. The states in the overlap region have gaps set by the matrix elements $V$ coupling them. Pairs of states localised nearby in space have gaps $V \gg \omega$, causing adiabatic following of the occupied lower band state~\cite{altshuler2010anderson_local_adiabatic,ved_nand_sond_local_adiabatic}. States with support far away from each other have matrix elements $V \ll \omega$, resulting in diabatic transition into the upper band, followed by another transition back into to the lower band. An even number of such crossings will lead to no transitions into the upper band. It is the states with $V = \mathcal{O} \left( \omega \right)$ that contribute to the inter band transitions between the instantaneous eigenstates (Fig.~\ref{fig:mechanisms} (b) inset). In the order of limits $L\to \infty$, $\omega \to 0$ each lower band state encounters a finite number of such crossings every period (Appendix A), with the probability of transition per period being $P_{\mathrm{exc}}\left(|\epsilon_0| < W-W_c,\omega, V=\mathcal{O}(\omega) \right)=\mathcal{O}(1)$. Non-adiabatic transitions between the states outside the overlapping bands have vanishing probability as $\omega \to 0$~\eqref{eq:trans_prob_non_ad}. 

\begin{figure}
    \centering
    \includegraphics[width=\linewidth]{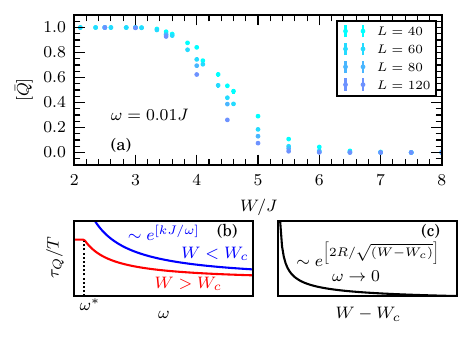}
    \caption{(a) Disorder averaged $\bar{Q}$ as a function of disorder strength $W$ at different system sizes. The transition at critical $W_c\approx 3.16 J$ (Appendix A) becomes sharper as $L \to \infty$ allowing for finite size scaling analysis in Fig.~\ref{fig:scaling_W}. \emph{Parameters}: $J=1.0$, $\Delta_0=1.5J$, $\delta_0=0.5J$, $\omega=0.01J$ and $[\bar{Q}]$ averaged over 200 disorder realizations. (b) Sketch showing the charge pumping decay time $\tau_Q$ as a function of $\omega$ above and below the critical disorder strength $W_c$ and (c) as a function of $W>W_c$ in the adiabatic limit $\omega \to 0$.}
    \label{fig:rare_r_physics}
\end{figure}

The full excitation rate due to the adiabatic scattering mechanism is calculated by adding up the contributions from the dominant transitions within the overlapping Lifshitz tails (Appendix A). Again, we treat subsequent transitions as incoherent in the thermodynamic limit. Initialising the system in one of the instantaneous bands, the pumping decays on a timescale 
\begin{equation}\label{eq:loc_time_W}
    \tau_Q \sim \frac{1}{\omega}\exp\left[ 2R\left( W-W_c \right)^{-1/2}\right].
\end{equation}
As the disorder strength approaches its critical value from above, in the adiabatic limit, the charge pumping decay timescale $\tau_Q$ diverges (Fig.~\ref{fig:rare_r_physics}(c)). 

As above, we consider the finite size scaling of the pump rate $\bar{Q}$ (Fig.~\ref{fig:rare_r_physics}(a)). The number of unit cells over which charge is transferred across the system is 
\begin{equation}\label{eq:loc_length_W}
    \xi_W=\frac{\tau_Q}{T} \sim \exp\left[ 2R\left( W-W_c \right)^{-1/2}\right].
\end{equation}
In Appendix B we show a scaling collapse consistent with the form in~\eqref{eq:scaling_form_omega}, where $\xi_{\omega}$ is replaced by $\xi_{W}$. Moreover, the flow with systems size in Fig.~\ref{fig:rare_r_physics} (a) indicates that the average pumped charge either remains fixed at $[\bar{Q}] = 1$, or flows to $[\bar{Q}] = 0$ with system size.

In the adiabatic limit, the quantised pumping decay time $\tau_Q/T$ becomes independent of $\omega$ when $W \geq W_c$ (Fig.~\ref{fig:rare_r_physics} (b)). However, the non-adiabatic mechanism remains present at finite $\omega$. The crossover frequency $\omega^*$ at which the adiabatic mechanism becomes dominant occurs when the decay timescales associated with the two breakdown mechanisms become comparable, equivalently when $\xi_{\omega} \approx \xi_W$. 

\emph{Discussion.}---Our result is the phase diagram of the Thouless pump at drive frequency $\omega \geq 0$ and disorder strengths $W \geq 0$ (Fig.~\ref{fig:phase_diag}(b)). Initialising the system in the experimentally accessible instantaneous ground state at integer filling, the phase diagram is as follows. In the adiabatic limit $(\omega \to 0)$ the charge pumping is quantised at weak disorder $W < W_c$. At any finite drive frequency and disorder strength $\omega, W>0$ the pump rate decays to zero in the steady state. The decay is exponentially long in the period of the drive $\tau_Q \sim \exp(\alpha/\omega)/\omega$ as $\omega \to 0$, where $\alpha$ is the band gap energy scale. On the other hand, the clean system pump rate $\bar{Q}$ assumes a finite, non-quantised, value at non-zero drive frequencies. At low frequencies, each state (with a well defined quasimomentum) belonging to one of the bands becomes coherently dressed with another state in the opposite pumping band. This delicate coherence effect vanishes with any random disorder, resulting in the pump rate $\bar{Q}$ decaying to zero for any finite disorder strength. 

The adiabatic mechanism prediction~\eqref{eq:loc_length_W} for the scaling form of $\xi_W$ at the point $(\omega, W)=(0, W_c)$ challenges a previous conjecture~\cite{PhysRevLett.123.266601_disorder} that the transition is in the integer quantum Hall (IQH) universality class~\cite{PhysRevLett.45.494_qhe,PhysRevB.23.5632_qhe,PhysRevLett.49.405_qhe,PhysRevB.63.054430_qhe,PhysRevLett.88.036401_qhe,
PhysRevLett.89.256601_iqhe_trans,PhysRevB.80.041304_iqhe_trans,PhysRevB.99.121301_iqhe_trans}. In Appendix B we show that the finite size scaling of the pump rate $\bar{Q}$ in Fig.~\ref{fig:rare_r_physics} (a) is consistent with both the IQH and Lifshitz tail driven transitions. Given that the IQH transition predicts the transverse conductivity $\sigma_{xy}=1/2$ at the critical point, one might expect that $[\bar{Q}]=1/2$ if the disordered Thouless pump transition was related to it. However, the finite size data in Fig.~\ref{fig:rare_r_physics} (a) indicates that there is no scale invariant point with $[\bar{Q}]= 1/2$ . The average pumped charge either remains fixed at $[\bar{Q}]=1$, or flows to $[\bar{Q}]=0$. Moreover, a recent study~\cite{grabarits2023floquetanderson} numerically extracted a power law scaling form for the pumping breakdown length $\xi_{\omega} \sim \omega^{-\theta}$, which disagrees with our prediction for the non-adiabatic breakdown of pumping, $\xi_{\omega} \sim \omega e^{kJ/\omega}$. Nonetheless, analysis of the data from Ref.~\cite{grabarits2023floquetanderson} indicates that it is consistent with our proposed scaling for sufficiently small omega (Appendix D).

The phase diagram can be experimentally probed using ultracold atoms and photonic platforms. Thouless pumping with random disorder has already been observed with waveguide arrays in the adiabatic limit~\cite{disordered_thouless_pump_photonics_exp}. Finite frequency effects, such as the decay in the pump rate can also be studied here. 

Ultracold atoms provide a natural platform to probe the stability of Thouless pumping to quasiperiodic disorder. Reference~\cite{disorder_thouless_pump_cold_atoms_exp} showed quantised pumping to be stable at weak disorder in the adiabatic limit. Many aspects of the phase diagrams of the disordered Thouless pump with quasiperiodic and random potentials are equivalent. However, in the localised phase of the Aubry-Andre model~\cite{aubry1980analyticity}, there are no Lifshitz tails at the band edges. Thus, the expected pump rate decay time scale $\tau_Q$ near the transition would be power law in $\left( W-W_c \right)$, in contrast to Eq.~\eqref{eq:loc_time_W}.

The Thouless pump can be mapped onto a two-dimensional tilted lattice model with the synthetic lattice construction~\cite{ PhysRevA.7.2203_synth_lat,  HO1983464_synth_lat,  VerdenyPuigMintert+2016+897+907_synth_lat, PhysRevX.7.041008_synth_latt_qubit, PhysRevB.97.134303_synth_lat, PhysRevB.98.220509_synth_lat, PhysRevB.99.064306_synth_lat_qubit, PhysRevLett.125.100601_synth_lat_qubit,PhysRevLett.126.106805_synth_lat}. In the pumping regime the eigenstates are spatially delocalised and current-carrying. Away from the adiabatic limit the counter-propagating states hybridise to give localised eigenstates of the lattice model (Appendix C). The localisation length of the synthetic eigenstates is exponentially large in the period of the drive---just as the length scale of charge pumping at finite frequency~\eqref{eq:omega_loc_length}.
The synthetic lattice brings out the similarities between the Thouless pump and other driven physical systems, such as the quasiperiodically driven qubit~\cite{PhysRevB.99.094311_qubit,PhysRevLett.125.160505_synth_lat_qubit, PhysRevResearch.2.043411_qubit, PhysRevB.99.064306_synth_lat_qubit, PhysRevX.7.041008_synth_latt_qubit, PhysRevLett.125.100601_synth_lat_qubit,PhysRevLett.128.183602_qubit, PhysRevB.108.134303_qubit} (Appendix C).

\emph{Acknowledgements}---The authors thank I. Esin, M. Kolodrubetz, C. Laumann for helpful discussions. We are grateful to M. Wauters for pointing us to Ref.~\cite{grabarits2023floquetanderson}. This work was supported by: NSF Grant No.DMR-1752759, and AFOSR Grant No. FA9550-20-1-0235 (D.V., D.L. and A.C.); and the NSF QII-TAQS program (P.C.). D.L. also acknowledges the support of the ``Laboratory for Physical Sciences''. Numerical work was performed on the BU Shared Computing Cluster, using \textsc{Quspin}~\cite{10.21468/SciPostPhys.2.1.003_quspin, 10.21468/SciPostPhys.7.2.020_quspin}.

\bibliography{ref.bib}

\begin{thebibliography}{97}%
\makeatletter
\providecommand \@ifxundefined [1]{%
 \@ifx{#1\undefined}
}%
\providecommand \@ifnum [1]{%
 \ifnum #1\expandafter \@firstoftwo
 \else \expandafter \@secondoftwo
 \fi
}%
\providecommand \@ifx [1]{%
 \ifx #1\expandafter \@firstoftwo
 \else \expandafter \@secondoftwo
 \fi
}%
\providecommand \natexlab [1]{#1}%
\providecommand \enquote  [1]{``#1''}%
\providecommand \bibnamefont  [1]{#1}%
\providecommand \bibfnamefont [1]{#1}%
\providecommand \citenamefont [1]{#1}%
\providecommand \href@noop [0]{\@secondoftwo}%
\providecommand \href [0]{\begingroup \@sanitize@url \@href}%
\providecommand \@href[1]{\@@startlink{#1}\@@href}%
\providecommand \@@href[1]{\endgroup#1\@@endlink}%
\providecommand \@sanitize@url [0]{\catcode `\\12\catcode `\$12\catcode `\&12\catcode `\#12\catcode `\^12\catcode `\_12\catcode `\%12\relax}%
\providecommand \@@startlink[1]{}%
\providecommand \@@endlink[0]{}%
\providecommand \url  [0]{\begingroup\@sanitize@url \@url }%
\providecommand \@url [1]{\endgroup\@href {#1}{\urlprefix }}%
\providecommand \urlprefix  [0]{URL }%
\providecommand \Eprint [0]{\href }%
\providecommand \doibase [0]{http://dx.doi.org/}%
\providecommand \selectlanguage [0]{\@gobble}%
\providecommand \bibinfo  [0]{\@secondoftwo}%
\providecommand \bibfield  [0]{\@secondoftwo}%
\providecommand \translation [1]{[#1]}%
\providecommand \BibitemOpen [0]{}%
\providecommand \bibitemStop [0]{}%
\providecommand \bibitemNoStop [0]{.\EOS\space}%
\providecommand \EOS [0]{\spacefactor3000\relax}%
\providecommand \BibitemShut  [1]{\csname bibitem#1\endcsname}%
\let\auto@bib@innerbib\@empty
\bibitem [{\citenamefont {Thouless}(1983)}]{PhysRevB.27.6083_tholuess_main}%
  \BibitemOpen
  \bibfield  {author} {\bibinfo {author} {\bibfnamefont {D.~J.}\ \bibnamefont {Thouless}},\ }\href {\doibase 10.1103/PhysRevB.27.6083} {\bibfield  {journal} {\bibinfo  {journal} {Phys. Rev. B}\ }\textbf {\bibinfo {volume} {27}},\ \bibinfo {pages} {6083} (\bibinfo {year} {1983})}\BibitemShut {NoStop}%
\bibitem [{\citenamefont {Citro}\ and\ \citenamefont {Aidelsburger}(2023)}]{citro_aidel_nature_rev}%
  \BibitemOpen
  \bibfield  {author} {\bibinfo {author} {\bibfnamefont {R.}~\bibnamefont {Citro}}\ and\ \bibinfo {author} {\bibfnamefont {M.}~\bibnamefont {Aidelsburger}},\ }\href {\doibase 10.1038/s42254-022-00545-0} {\bibfield  {journal} {\bibinfo  {journal} {Nature Reviews Physics}\ }\textbf {\bibinfo {volume} {5}},\ \bibinfo {pages} {87} (\bibinfo {year} {2023})}\BibitemShut {NoStop}%
\bibitem [{\citenamefont {Nakajima}\ \emph {et~al.}(2016)\citenamefont {Nakajima}, \citenamefont {Tomita}, \citenamefont {Taie}, \citenamefont {Ichinose}, \citenamefont {Ozawa}, \citenamefont {Wang}, \citenamefont {Troyer},\ and\ \citenamefont {Takahashi}}]{thouless_pump_cold_atoms_exp_fermions}%
  \BibitemOpen
  \bibfield  {author} {\bibinfo {author} {\bibfnamefont {S.}~\bibnamefont {Nakajima}}, \bibinfo {author} {\bibfnamefont {T.}~\bibnamefont {Tomita}}, \bibinfo {author} {\bibfnamefont {S.}~\bibnamefont {Taie}}, \bibinfo {author} {\bibfnamefont {T.}~\bibnamefont {Ichinose}}, \bibinfo {author} {\bibfnamefont {H.}~\bibnamefont {Ozawa}}, \bibinfo {author} {\bibfnamefont {L.}~\bibnamefont {Wang}}, \bibinfo {author} {\bibfnamefont {M.}~\bibnamefont {Troyer}}, \ and\ \bibinfo {author} {\bibfnamefont {Y.}~\bibnamefont {Takahashi}},\ }\href {\doibase 10.1038/nphys3622} {\bibfield  {journal} {\bibinfo  {journal} {Nature Physics}\ }\textbf {\bibinfo {volume} {12}},\ \bibinfo {pages} {296} (\bibinfo {year} {2016})}\BibitemShut {NoStop}%
\bibitem [{\citenamefont {Lohse}\ \emph {et~al.}(2016)\citenamefont {Lohse}, \citenamefont {Schweizer}, \citenamefont {Zilberberg}, \citenamefont {Aidelsburger},\ and\ \citenamefont {Bloch}}]{thouless_pump_cold_atoms_exp_bosons}%
  \BibitemOpen
  \bibfield  {author} {\bibinfo {author} {\bibfnamefont {M.}~\bibnamefont {Lohse}}, \bibinfo {author} {\bibfnamefont {C.}~\bibnamefont {Schweizer}}, \bibinfo {author} {\bibfnamefont {O.}~\bibnamefont {Zilberberg}}, \bibinfo {author} {\bibfnamefont {M.}~\bibnamefont {Aidelsburger}}, \ and\ \bibinfo {author} {\bibfnamefont {I.}~\bibnamefont {Bloch}},\ }\href {\doibase 10.1038/nphys3584} {\bibfield  {journal} {\bibinfo  {journal} {Nature Physics}\ }\textbf {\bibinfo {volume} {12}},\ \bibinfo {pages} {350} (\bibinfo {year} {2016})}\BibitemShut {NoStop}%
\bibitem [{\citenamefont {Lu}\ \emph {et~al.}(2016)\citenamefont {Lu}, \citenamefont {Schemmer}, \citenamefont {Aycock}, \citenamefont {Genkina}, \citenamefont {Sugawa},\ and\ \citenamefont {Spielman}}]{PhysRevLett.116.200402_exp_cold_atoms}%
  \BibitemOpen
  \bibfield  {author} {\bibinfo {author} {\bibfnamefont {H.-I.}\ \bibnamefont {Lu}}, \bibinfo {author} {\bibfnamefont {M.}~\bibnamefont {Schemmer}}, \bibinfo {author} {\bibfnamefont {L.~M.}\ \bibnamefont {Aycock}}, \bibinfo {author} {\bibfnamefont {D.}~\bibnamefont {Genkina}}, \bibinfo {author} {\bibfnamefont {S.}~\bibnamefont {Sugawa}}, \ and\ \bibinfo {author} {\bibfnamefont {I.~B.}\ \bibnamefont {Spielman}},\ }\href {\doibase 10.1103/PhysRevLett.116.200402} {\bibfield  {journal} {\bibinfo  {journal} {Phys. Rev. Lett.}\ }\textbf {\bibinfo {volume} {116}},\ \bibinfo {pages} {200402} (\bibinfo {year} {2016})}\BibitemShut {NoStop}%
\bibitem [{\citenamefont {Nakajima}\ \emph {et~al.}(2021)\citenamefont {Nakajima}, \citenamefont {Takei}, \citenamefont {Sakuma}, \citenamefont {Kuno}, \citenamefont {Marra},\ and\ \citenamefont {Takahashi}}]{disorder_thouless_pump_cold_atoms_exp}%
  \BibitemOpen
  \bibfield  {author} {\bibinfo {author} {\bibfnamefont {S.}~\bibnamefont {Nakajima}}, \bibinfo {author} {\bibfnamefont {N.}~\bibnamefont {Takei}}, \bibinfo {author} {\bibfnamefont {K.}~\bibnamefont {Sakuma}}, \bibinfo {author} {\bibfnamefont {Y.}~\bibnamefont {Kuno}}, \bibinfo {author} {\bibfnamefont {P.}~\bibnamefont {Marra}}, \ and\ \bibinfo {author} {\bibfnamefont {Y.}~\bibnamefont {Takahashi}},\ }\href {\doibase 10.1038/s41567-021-01229-9} {\bibfield  {journal} {\bibinfo  {journal} {Nature Physics}\ }\textbf {\bibinfo {volume} {17}},\ \bibinfo {pages} {844} (\bibinfo {year} {2021})}\BibitemShut {NoStop}%
\bibitem [{\citenamefont {Minguzzi}\ \emph {et~al.}(2022)\citenamefont {Minguzzi}, \citenamefont {Zhu}, \citenamefont {Sandholzer}, \citenamefont {Walter}, \citenamefont {Viebahn},\ and\ \citenamefont {Esslinger}}]{PhysRevLett.129.053201_ultracold_atoms}%
  \BibitemOpen
  \bibfield  {author} {\bibinfo {author} {\bibfnamefont {J.}~\bibnamefont {Minguzzi}}, \bibinfo {author} {\bibfnamefont {Z.}~\bibnamefont {Zhu}}, \bibinfo {author} {\bibfnamefont {K.}~\bibnamefont {Sandholzer}}, \bibinfo {author} {\bibfnamefont {A.-S.}\ \bibnamefont {Walter}}, \bibinfo {author} {\bibfnamefont {K.}~\bibnamefont {Viebahn}}, \ and\ \bibinfo {author} {\bibfnamefont {T.}~\bibnamefont {Esslinger}},\ }\href {\doibase 10.1103/PhysRevLett.129.053201} {\bibfield  {journal} {\bibinfo  {journal} {Phys. Rev. Lett.}\ }\textbf {\bibinfo {volume} {129}},\ \bibinfo {pages} {053201} (\bibinfo {year} {2022})}\BibitemShut {NoStop}%
\bibitem [{\citenamefont {Walter}\ \emph {et~al.}(2023)\citenamefont {Walter}, \citenamefont {Zhu}, \citenamefont {G{\"a}chter}, \citenamefont {Minguzzi}, \citenamefont {Roschinski}, \citenamefont {Sandholzer}, \citenamefont {Viebahn},\ and\ \citenamefont {Esslinger}}]{interactions_exp_cold_atoms}%
  \BibitemOpen
  \bibfield  {author} {\bibinfo {author} {\bibfnamefont {A.-S.}\ \bibnamefont {Walter}}, \bibinfo {author} {\bibfnamefont {Z.}~\bibnamefont {Zhu}}, \bibinfo {author} {\bibfnamefont {M.}~\bibnamefont {G{\"a}chter}}, \bibinfo {author} {\bibfnamefont {J.}~\bibnamefont {Minguzzi}}, \bibinfo {author} {\bibfnamefont {S.}~\bibnamefont {Roschinski}}, \bibinfo {author} {\bibfnamefont {K.}~\bibnamefont {Sandholzer}}, \bibinfo {author} {\bibfnamefont {K.}~\bibnamefont {Viebahn}}, \ and\ \bibinfo {author} {\bibfnamefont {T.}~\bibnamefont {Esslinger}},\ }\href {\doibase 10.1038/s41567-023-02145-w} {\bibfield  {journal} {\bibinfo  {journal} {Nature Physics}\ }\textbf {\bibinfo {volume} {19}},\ \bibinfo {pages} {1471} (\bibinfo {year} {2023})}\BibitemShut {NoStop}%
\bibitem [{\citenamefont {Kraus}\ \emph {et~al.}(2012)\citenamefont {Kraus}, \citenamefont {Lahini}, \citenamefont {Ringel}, \citenamefont {Verbin},\ and\ \citenamefont {Zilberberg}}]{PhysRevLett.109.106402_photo_exp}%
  \BibitemOpen
  \bibfield  {author} {\bibinfo {author} {\bibfnamefont {Y.~E.}\ \bibnamefont {Kraus}}, \bibinfo {author} {\bibfnamefont {Y.}~\bibnamefont {Lahini}}, \bibinfo {author} {\bibfnamefont {Z.}~\bibnamefont {Ringel}}, \bibinfo {author} {\bibfnamefont {M.}~\bibnamefont {Verbin}}, \ and\ \bibinfo {author} {\bibfnamefont {O.}~\bibnamefont {Zilberberg}},\ }\href {\doibase 10.1103/PhysRevLett.109.106402} {\bibfield  {journal} {\bibinfo  {journal} {Phys. Rev. Lett.}\ }\textbf {\bibinfo {volume} {109}},\ \bibinfo {pages} {106402} (\bibinfo {year} {2012})}\BibitemShut {NoStop}%
\bibitem [{\citenamefont {Verbin}\ \emph {et~al.}(2015)\citenamefont {Verbin}, \citenamefont {Zilberberg}, \citenamefont {Lahini}, \citenamefont {Kraus},\ and\ \citenamefont {Silberberg}}]{PhysRevB.91.064201_photo_exp}%
  \BibitemOpen
  \bibfield  {author} {\bibinfo {author} {\bibfnamefont {M.}~\bibnamefont {Verbin}}, \bibinfo {author} {\bibfnamefont {O.}~\bibnamefont {Zilberberg}}, \bibinfo {author} {\bibfnamefont {Y.}~\bibnamefont {Lahini}}, \bibinfo {author} {\bibfnamefont {Y.~E.}\ \bibnamefont {Kraus}}, \ and\ \bibinfo {author} {\bibfnamefont {Y.}~\bibnamefont {Silberberg}},\ }\href {\doibase 10.1103/PhysRevB.91.064201} {\bibfield  {journal} {\bibinfo  {journal} {Phys. Rev. B}\ }\textbf {\bibinfo {volume} {91}},\ \bibinfo {pages} {064201} (\bibinfo {year} {2015})}\BibitemShut {NoStop}%
\bibitem [{\citenamefont {Ke}\ \emph {et~al.}(2016)\citenamefont {Ke}, \citenamefont {Qin}, \citenamefont {Mei}, \citenamefont {Zhong}, \citenamefont {Kivshar},\ and\ \citenamefont {Lee}}]{Ke_2016_exp_photonics}%
  \BibitemOpen
  \bibfield  {author} {\bibinfo {author} {\bibfnamefont {Y.}~\bibnamefont {Ke}}, \bibinfo {author} {\bibfnamefont {X.}~\bibnamefont {Qin}}, \bibinfo {author} {\bibfnamefont {F.}~\bibnamefont {Mei}}, \bibinfo {author} {\bibfnamefont {H.}~\bibnamefont {Zhong}}, \bibinfo {author} {\bibfnamefont {Y.~S.}\ \bibnamefont {Kivshar}}, \ and\ \bibinfo {author} {\bibfnamefont {C.}~\bibnamefont {Lee}},\ }\href {\doibase 10.1002/lpor.201600119} {\bibfield  {journal} {\bibinfo  {journal} {Laser \& Photonics Reviews}\ }\textbf {\bibinfo {volume} {10}},\ \bibinfo {pages} {995} (\bibinfo {year} {2016})}\BibitemShut {NoStop}%
\bibitem [{\citenamefont {Tangpanitanon}\ \emph {et~al.}(2016)\citenamefont {Tangpanitanon}, \citenamefont {Bastidas}, \citenamefont {Al-Assam}, \citenamefont {Roushan}, \citenamefont {Jaksch},\ and\ \citenamefont {Angelakis}}]{PhysRevLett.117.213603_interactions_exp_photonics}%
  \BibitemOpen
  \bibfield  {author} {\bibinfo {author} {\bibfnamefont {J.}~\bibnamefont {Tangpanitanon}}, \bibinfo {author} {\bibfnamefont {V.~M.}\ \bibnamefont {Bastidas}}, \bibinfo {author} {\bibfnamefont {S.}~\bibnamefont {Al-Assam}}, \bibinfo {author} {\bibfnamefont {P.}~\bibnamefont {Roushan}}, \bibinfo {author} {\bibfnamefont {D.}~\bibnamefont {Jaksch}}, \ and\ \bibinfo {author} {\bibfnamefont {D.~G.}\ \bibnamefont {Angelakis}},\ }\href {\doibase 10.1103/PhysRevLett.117.213603} {\bibfield  {journal} {\bibinfo  {journal} {Phys. Rev. Lett.}\ }\textbf {\bibinfo {volume} {117}},\ \bibinfo {pages} {213603} (\bibinfo {year} {2016})}\BibitemShut {NoStop}%
\bibitem [{\citenamefont {Fedorova}\ \emph {et~al.}(2020)\citenamefont {Fedorova}, \citenamefont {Qiu}, \citenamefont {Linden},\ and\ \citenamefont {Kroha}}]{dissipation_exp_photonics1}%
  \BibitemOpen
  \bibfield  {author} {\bibinfo {author} {\bibfnamefont {Z.}~\bibnamefont {Fedorova}}, \bibinfo {author} {\bibfnamefont {H.}~\bibnamefont {Qiu}}, \bibinfo {author} {\bibfnamefont {S.}~\bibnamefont {Linden}}, \ and\ \bibinfo {author} {\bibfnamefont {J.}~\bibnamefont {Kroha}},\ }\href {\doibase 10.1038/s41467-020-17510-z} {\bibfield  {journal} {\bibinfo  {journal} {Nature Communications}\ }\textbf {\bibinfo {volume} {11}},\ \bibinfo {pages} {3758} (\bibinfo {year} {2020})}\BibitemShut {NoStop}%
\bibitem [{\citenamefont {Cerjan}\ \emph {et~al.}(2020)\citenamefont {Cerjan}, \citenamefont {Wang}, \citenamefont {Huang}, \citenamefont {Chen},\ and\ \citenamefont {Rechtsman}}]{disordered_thouless_pump_photonics_exp}%
  \BibitemOpen
  \bibfield  {author} {\bibinfo {author} {\bibfnamefont {A.}~\bibnamefont {Cerjan}}, \bibinfo {author} {\bibfnamefont {M.}~\bibnamefont {Wang}}, \bibinfo {author} {\bibfnamefont {S.}~\bibnamefont {Huang}}, \bibinfo {author} {\bibfnamefont {K.~P.}\ \bibnamefont {Chen}}, \ and\ \bibinfo {author} {\bibfnamefont {M.~C.}\ \bibnamefont {Rechtsman}},\ }\href {\doibase 10.1038/s41377-020-00408-2} {\bibfield  {journal} {\bibinfo  {journal} {Light: Science \& Applications}\ }\textbf {\bibinfo {volume} {9}},\ \bibinfo {pages} {178} (\bibinfo {year} {2020})}\BibitemShut {NoStop}%
\bibitem [{\citenamefont {J{\"u}rgensen}\ \emph {et~al.}(2021)\citenamefont {J{\"u}rgensen}, \citenamefont {Mukherjee},\ and\ \citenamefont {Rechtsman}}]{interactions_exp_photonics}%
  \BibitemOpen
  \bibfield  {author} {\bibinfo {author} {\bibfnamefont {M.}~\bibnamefont {J{\"u}rgensen}}, \bibinfo {author} {\bibfnamefont {S.}~\bibnamefont {Mukherjee}}, \ and\ \bibinfo {author} {\bibfnamefont {M.~C.}\ \bibnamefont {Rechtsman}},\ }\href {\doibase 10.1038/s41586-021-03688-9} {\bibfield  {journal} {\bibinfo  {journal} {Nature}\ }\textbf {\bibinfo {volume} {596}},\ \bibinfo {pages} {63} (\bibinfo {year} {2021})}\BibitemShut {NoStop}%
\bibitem [{\citenamefont {Dreon}\ \emph {et~al.}(2022)\citenamefont {Dreon}, \citenamefont {Baumg{\"a}rtner}, \citenamefont {Li}, \citenamefont {Hertlein}, \citenamefont {Esslinger},\ and\ \citenamefont {Donner}}]{dissipation_exp_photonics2}%
  \BibitemOpen
  \bibfield  {author} {\bibinfo {author} {\bibfnamefont {D.}~\bibnamefont {Dreon}}, \bibinfo {author} {\bibfnamefont {A.}~\bibnamefont {Baumg{\"a}rtner}}, \bibinfo {author} {\bibfnamefont {X.}~\bibnamefont {Li}}, \bibinfo {author} {\bibfnamefont {S.}~\bibnamefont {Hertlein}}, \bibinfo {author} {\bibfnamefont {T.}~\bibnamefont {Esslinger}}, \ and\ \bibinfo {author} {\bibfnamefont {T.}~\bibnamefont {Donner}},\ }\href {\doibase 10.1038/s41586-022-04970-0} {\bibfield  {journal} {\bibinfo  {journal} {Nature}\ }\textbf {\bibinfo {volume} {608}},\ \bibinfo {pages} {494} (\bibinfo {year} {2022})}\BibitemShut {NoStop}%
\bibitem [{\citenamefont {Niu}\ and\ \citenamefont {Thouless}(1984)}]{niu1984quantised}%
  \BibitemOpen
  \bibfield  {author} {\bibinfo {author} {\bibfnamefont {Q.}~\bibnamefont {Niu}}\ and\ \bibinfo {author} {\bibfnamefont {D.}~\bibnamefont {Thouless}},\ }\href@noop {} {\bibfield  {journal} {\bibinfo  {journal} {Journal of Physics A: Mathematical and General}\ }\textbf {\bibinfo {volume} {17}},\ \bibinfo {pages} {2453} (\bibinfo {year} {1984})}\BibitemShut {NoStop}%
\bibitem [{\citenamefont {Niu}\ and\ \citenamefont {Thouless}(1987)}]{PhysRevB.35.2188_exponentially_small_corr}%
  \BibitemOpen
  \bibfield  {author} {\bibinfo {author} {\bibfnamefont {Q.}~\bibnamefont {Niu}}\ and\ \bibinfo {author} {\bibfnamefont {D.~J.}\ \bibnamefont {Thouless}},\ }\href {\doibase 10.1103/PhysRevB.35.2188} {\bibfield  {journal} {\bibinfo  {journal} {Phys. Rev. B}\ }\textbf {\bibinfo {volume} {35}},\ \bibinfo {pages} {2188} (\bibinfo {year} {1987})}\BibitemShut {NoStop}%
\bibitem [{\citenamefont {Thouless}(1998)}]{thouless1998topological_exponentially_small_corr}%
  \BibitemOpen
  \bibfield  {author} {\bibinfo {author} {\bibfnamefont {D.}~\bibnamefont {Thouless}},\ }\href@noop {} {\emph {\bibinfo {title} {Topological quantum numbers in nonrelativistic physics}}}\ (\bibinfo  {publisher} {World Scientific},\ \bibinfo {year} {1998})\BibitemShut {NoStop}%
\bibitem [{\citenamefont {Qin}\ and\ \citenamefont {Guo}(2016)}]{QIN20162317_disorder}%
  \BibitemOpen
  \bibfield  {author} {\bibinfo {author} {\bibfnamefont {J.}~\bibnamefont {Qin}}\ and\ \bibinfo {author} {\bibfnamefont {H.}~\bibnamefont {Guo}},\ }\href {\doibase https://doi.org/10.1016/j.physleta.2016.05.014} {\bibfield  {journal} {\bibinfo  {journal} {Physics Letters A}\ }\textbf {\bibinfo {volume} {380}},\ \bibinfo {pages} {2317} (\bibinfo {year} {2016})}\BibitemShut {NoStop}%
\bibitem [{\citenamefont {Wang}\ and\ \citenamefont {Song}(2019)}]{PhysRevB.100.184304_disorder}%
  \BibitemOpen
  \bibfield  {author} {\bibinfo {author} {\bibfnamefont {R.}~\bibnamefont {Wang}}\ and\ \bibinfo {author} {\bibfnamefont {Z.}~\bibnamefont {Song}},\ }\href {\doibase 10.1103/PhysRevB.100.184304} {\bibfield  {journal} {\bibinfo  {journal} {Phys. Rev. B}\ }\textbf {\bibinfo {volume} {100}},\ \bibinfo {pages} {184304} (\bibinfo {year} {2019})}\BibitemShut {NoStop}%
\bibitem [{\citenamefont {Wauters}\ \emph {et~al.}(2019)\citenamefont {Wauters}, \citenamefont {Russomanno}, \citenamefont {Citro}, \citenamefont {Santoro},\ and\ \citenamefont {Privitera}}]{PhysRevLett.123.266601_disorder}%
  \BibitemOpen
  \bibfield  {author} {\bibinfo {author} {\bibfnamefont {M.~M.}\ \bibnamefont {Wauters}}, \bibinfo {author} {\bibfnamefont {A.}~\bibnamefont {Russomanno}}, \bibinfo {author} {\bibfnamefont {R.}~\bibnamefont {Citro}}, \bibinfo {author} {\bibfnamefont {G.~E.}\ \bibnamefont {Santoro}}, \ and\ \bibinfo {author} {\bibfnamefont {L.}~\bibnamefont {Privitera}},\ }\href {\doibase 10.1103/PhysRevLett.123.266601} {\bibfield  {journal} {\bibinfo  {journal} {Phys. Rev. Lett.}\ }\textbf {\bibinfo {volume} {123}},\ \bibinfo {pages} {266601} (\bibinfo {year} {2019})}\BibitemShut {NoStop}%
\bibitem [{\citenamefont {Ippoliti}\ and\ \citenamefont {Bhatt}(2020)}]{PhysRevLett.124.086602_disorder}%
  \BibitemOpen
  \bibfield  {author} {\bibinfo {author} {\bibfnamefont {M.}~\bibnamefont {Ippoliti}}\ and\ \bibinfo {author} {\bibfnamefont {R.~N.}\ \bibnamefont {Bhatt}},\ }\href {\doibase 10.1103/PhysRevLett.124.086602} {\bibfield  {journal} {\bibinfo  {journal} {Phys. Rev. Lett.}\ }\textbf {\bibinfo {volume} {124}},\ \bibinfo {pages} {086602} (\bibinfo {year} {2020})}\BibitemShut {NoStop}%
\bibitem [{\citenamefont {Marra}\ and\ \citenamefont {Nitta}(2020)}]{PhysRevResearch.2.042035_disorder}%
  \BibitemOpen
  \bibfield  {author} {\bibinfo {author} {\bibfnamefont {P.}~\bibnamefont {Marra}}\ and\ \bibinfo {author} {\bibfnamefont {M.}~\bibnamefont {Nitta}},\ }\href {\doibase 10.1103/PhysRevResearch.2.042035} {\bibfield  {journal} {\bibinfo  {journal} {Phys. Rev. Res.}\ }\textbf {\bibinfo {volume} {2}},\ \bibinfo {pages} {042035} (\bibinfo {year} {2020})}\BibitemShut {NoStop}%
\bibitem [{\citenamefont {Hu}\ \emph {et~al.}(2020)\citenamefont {Hu}, \citenamefont {Ke},\ and\ \citenamefont {Lee}}]{PhysRevA.101.052323_disorder}%
  \BibitemOpen
  \bibfield  {author} {\bibinfo {author} {\bibfnamefont {S.}~\bibnamefont {Hu}}, \bibinfo {author} {\bibfnamefont {Y.}~\bibnamefont {Ke}}, \ and\ \bibinfo {author} {\bibfnamefont {C.}~\bibnamefont {Lee}},\ }\href {\doibase 10.1103/PhysRevA.101.052323} {\bibfield  {journal} {\bibinfo  {journal} {Phys. Rev. A}\ }\textbf {\bibinfo {volume} {101}},\ \bibinfo {pages} {052323} (\bibinfo {year} {2020})}\BibitemShut {NoStop}%
\bibitem [{\citenamefont {Hayward}\ \emph {et~al.}(2021)\citenamefont {Hayward}, \citenamefont {Bertok}, \citenamefont {Schneider},\ and\ \citenamefont {Heidrich-Meisner}}]{PhysRevA.103.043310_disorder}%
  \BibitemOpen
  \bibfield  {author} {\bibinfo {author} {\bibfnamefont {A.~L.~C.}\ \bibnamefont {Hayward}}, \bibinfo {author} {\bibfnamefont {E.}~\bibnamefont {Bertok}}, \bibinfo {author} {\bibfnamefont {U.}~\bibnamefont {Schneider}}, \ and\ \bibinfo {author} {\bibfnamefont {F.}~\bibnamefont {Heidrich-Meisner}},\ }\href {\doibase 10.1103/PhysRevA.103.043310} {\bibfield  {journal} {\bibinfo  {journal} {Phys. Rev. A}\ }\textbf {\bibinfo {volume} {103}},\ \bibinfo {pages} {043310} (\bibinfo {year} {2021})}\BibitemShut {NoStop}%
\bibitem [{\citenamefont {Wu}\ \emph {et~al.}(2022)\citenamefont {Wu}, \citenamefont {Tang}, \citenamefont {Zhang},\ and\ \citenamefont {Zhang}}]{PhysRevA.106.L051301_disorder}%
  \BibitemOpen
  \bibfield  {author} {\bibinfo {author} {\bibfnamefont {Y.-P.}\ \bibnamefont {Wu}}, \bibinfo {author} {\bibfnamefont {L.-Z.}\ \bibnamefont {Tang}}, \bibinfo {author} {\bibfnamefont {G.-Q.}\ \bibnamefont {Zhang}}, \ and\ \bibinfo {author} {\bibfnamefont {D.-W.}\ \bibnamefont {Zhang}},\ }\href {\doibase 10.1103/PhysRevA.106.L051301} {\bibfield  {journal} {\bibinfo  {journal} {Phys. Rev. A}\ }\textbf {\bibinfo {volume} {106}},\ \bibinfo {pages} {L051301} (\bibinfo {year} {2022})}\BibitemShut {NoStop}%
\bibitem [{\citenamefont {Grabarits}\ \emph {et~al.}(2023)\citenamefont {Grabarits}, \citenamefont {Takács}, \citenamefont {Fulga},\ and\ \citenamefont {Asbóth}}]{grabarits2023floquetanderson}%
  \BibitemOpen
  \bibfield  {author} {\bibinfo {author} {\bibfnamefont {A.}~\bibnamefont {Grabarits}}, \bibinfo {author} {\bibfnamefont {A.}~\bibnamefont {Takács}}, \bibinfo {author} {\bibfnamefont {I.~C.}\ \bibnamefont {Fulga}}, \ and\ \bibinfo {author} {\bibfnamefont {J.~K.}\ \bibnamefont {Asbóth}},\ }\href@noop {} {\enquote {\bibinfo {title} {Floquet-anderson localization in the thouless pump and how to avoid it},}\ } (\bibinfo {year} {2023}),\ \Eprint {http://arxiv.org/abs/2309.12882} {arXiv:2309.12882 [cond-mat.dis-nn]} \BibitemShut {NoStop}%
\bibitem [{\citenamefont {Citro}\ \emph {et~al.}(2003)\citenamefont {Citro}, \citenamefont {Andrei},\ and\ \citenamefont {Niu}}]{PhysRevB.68.165312_interactions}%
  \BibitemOpen
  \bibfield  {author} {\bibinfo {author} {\bibfnamefont {R.}~\bibnamefont {Citro}}, \bibinfo {author} {\bibfnamefont {N.}~\bibnamefont {Andrei}}, \ and\ \bibinfo {author} {\bibfnamefont {Q.}~\bibnamefont {Niu}},\ }\href {\doibase 10.1103/PhysRevB.68.165312} {\bibfield  {journal} {\bibinfo  {journal} {Phys. Rev. B}\ }\textbf {\bibinfo {volume} {68}},\ \bibinfo {pages} {165312} (\bibinfo {year} {2003})}\BibitemShut {NoStop}%
\bibitem [{\citenamefont {Berg}\ \emph {et~al.}(2011)\citenamefont {Berg}, \citenamefont {Levin},\ and\ \citenamefont {Altman}}]{PhysRevLett.106.110405_interactions}%
  \BibitemOpen
  \bibfield  {author} {\bibinfo {author} {\bibfnamefont {E.}~\bibnamefont {Berg}}, \bibinfo {author} {\bibfnamefont {M.}~\bibnamefont {Levin}}, \ and\ \bibinfo {author} {\bibfnamefont {E.}~\bibnamefont {Altman}},\ }\href {\doibase 10.1103/PhysRevLett.106.110405} {\bibfield  {journal} {\bibinfo  {journal} {Phys. Rev. Lett.}\ }\textbf {\bibinfo {volume} {106}},\ \bibinfo {pages} {110405} (\bibinfo {year} {2011})}\BibitemShut {NoStop}%
\bibitem [{\citenamefont {Qian}\ \emph {et~al.}(2011)\citenamefont {Qian}, \citenamefont {Gong},\ and\ \citenamefont {Zhang}}]{PhysRevA.84.013608_interactions}%
  \BibitemOpen
  \bibfield  {author} {\bibinfo {author} {\bibfnamefont {Y.}~\bibnamefont {Qian}}, \bibinfo {author} {\bibfnamefont {M.}~\bibnamefont {Gong}}, \ and\ \bibinfo {author} {\bibfnamefont {C.}~\bibnamefont {Zhang}},\ }\href {\doibase 10.1103/PhysRevA.84.013608} {\bibfield  {journal} {\bibinfo  {journal} {Phys. Rev. A}\ }\textbf {\bibinfo {volume} {84}},\ \bibinfo {pages} {013608} (\bibinfo {year} {2011})}\BibitemShut {NoStop}%
\bibitem [{\citenamefont {Russomanno}\ \emph {et~al.}(2012)\citenamefont {Russomanno}, \citenamefont {Silva},\ and\ \citenamefont {Santoro}}]{PhysRevLett.109.257201_interactions}%
  \BibitemOpen
  \bibfield  {author} {\bibinfo {author} {\bibfnamefont {A.}~\bibnamefont {Russomanno}}, \bibinfo {author} {\bibfnamefont {A.}~\bibnamefont {Silva}}, \ and\ \bibinfo {author} {\bibfnamefont {G.~E.}\ \bibnamefont {Santoro}},\ }\href {\doibase 10.1103/PhysRevLett.109.257201} {\bibfield  {journal} {\bibinfo  {journal} {Phys. Rev. Lett.}\ }\textbf {\bibinfo {volume} {109}},\ \bibinfo {pages} {257201} (\bibinfo {year} {2012})}\BibitemShut {NoStop}%
\bibitem [{\citenamefont {Grusdt}\ and\ \citenamefont {H\"oning}(2014)}]{PhysRevA.90.053623_interactions}%
  \BibitemOpen
  \bibfield  {author} {\bibinfo {author} {\bibfnamefont {F.}~\bibnamefont {Grusdt}}\ and\ \bibinfo {author} {\bibfnamefont {M.}~\bibnamefont {H\"oning}},\ }\href {\doibase 10.1103/PhysRevA.90.053623} {\bibfield  {journal} {\bibinfo  {journal} {Phys. Rev. A}\ }\textbf {\bibinfo {volume} {90}},\ \bibinfo {pages} {053623} (\bibinfo {year} {2014})}\BibitemShut {NoStop}%
\bibitem [{\citenamefont {Zeng}\ \emph {et~al.}(2016)\citenamefont {Zeng}, \citenamefont {Zhu},\ and\ \citenamefont {Sheng}}]{PhysRevB.94.235139_interactions}%
  \BibitemOpen
  \bibfield  {author} {\bibinfo {author} {\bibfnamefont {T.-S.}\ \bibnamefont {Zeng}}, \bibinfo {author} {\bibfnamefont {W.}~\bibnamefont {Zhu}}, \ and\ \bibinfo {author} {\bibfnamefont {D.~N.}\ \bibnamefont {Sheng}},\ }\href {\doibase 10.1103/PhysRevB.94.235139} {\bibfield  {journal} {\bibinfo  {journal} {Phys. Rev. B}\ }\textbf {\bibinfo {volume} {94}},\ \bibinfo {pages} {235139} (\bibinfo {year} {2016})}\BibitemShut {NoStop}%
\bibitem [{\citenamefont {Lindner}\ \emph {et~al.}(2017)\citenamefont {Lindner}, \citenamefont {Berg},\ and\ \citenamefont {Rudner}}]{PhysRevX.7.011018_interactions}%
  \BibitemOpen
  \bibfield  {author} {\bibinfo {author} {\bibfnamefont {N.~H.}\ \bibnamefont {Lindner}}, \bibinfo {author} {\bibfnamefont {E.}~\bibnamefont {Berg}}, \ and\ \bibinfo {author} {\bibfnamefont {M.~S.}\ \bibnamefont {Rudner}},\ }\href {\doibase 10.1103/PhysRevX.7.011018} {\bibfield  {journal} {\bibinfo  {journal} {Phys. Rev. X}\ }\textbf {\bibinfo {volume} {7}},\ \bibinfo {pages} {011018} (\bibinfo {year} {2017})}\BibitemShut {NoStop}%
\bibitem [{\citenamefont {Hayward}\ \emph {et~al.}(2018)\citenamefont {Hayward}, \citenamefont {Schweizer}, \citenamefont {Lohse}, \citenamefont {Aidelsburger},\ and\ \citenamefont {Heidrich-Meisner}}]{PhysRevB.98.245148_interactions}%
  \BibitemOpen
  \bibfield  {author} {\bibinfo {author} {\bibfnamefont {A.}~\bibnamefont {Hayward}}, \bibinfo {author} {\bibfnamefont {C.}~\bibnamefont {Schweizer}}, \bibinfo {author} {\bibfnamefont {M.}~\bibnamefont {Lohse}}, \bibinfo {author} {\bibfnamefont {M.}~\bibnamefont {Aidelsburger}}, \ and\ \bibinfo {author} {\bibfnamefont {F.}~\bibnamefont {Heidrich-Meisner}},\ }\href {\doibase 10.1103/PhysRevB.98.245148} {\bibfield  {journal} {\bibinfo  {journal} {Phys. Rev. B}\ }\textbf {\bibinfo {volume} {98}},\ \bibinfo {pages} {245148} (\bibinfo {year} {2018})}\BibitemShut {NoStop}%
\bibitem [{\citenamefont {Stenzel}\ \emph {et~al.}(2019)\citenamefont {Stenzel}, \citenamefont {Hayward}, \citenamefont {Hubig}, \citenamefont {Schollw\"ock},\ and\ \citenamefont {Heidrich-Meisner}}]{PhysRevA.99.053614_interactions}%
  \BibitemOpen
  \bibfield  {author} {\bibinfo {author} {\bibfnamefont {L.}~\bibnamefont {Stenzel}}, \bibinfo {author} {\bibfnamefont {A.~L.~C.}\ \bibnamefont {Hayward}}, \bibinfo {author} {\bibfnamefont {C.}~\bibnamefont {Hubig}}, \bibinfo {author} {\bibfnamefont {U.}~\bibnamefont {Schollw\"ock}}, \ and\ \bibinfo {author} {\bibfnamefont {F.}~\bibnamefont {Heidrich-Meisner}},\ }\href {\doibase 10.1103/PhysRevA.99.053614} {\bibfield  {journal} {\bibinfo  {journal} {Phys. Rev. A}\ }\textbf {\bibinfo {volume} {99}},\ \bibinfo {pages} {053614} (\bibinfo {year} {2019})}\BibitemShut {NoStop}%
\bibitem [{\citenamefont {Greschner}\ \emph {et~al.}(2020)\citenamefont {Greschner}, \citenamefont {Mondal},\ and\ \citenamefont {Mishra}}]{PhysRevA.101.053630_interactions}%
  \BibitemOpen
  \bibfield  {author} {\bibinfo {author} {\bibfnamefont {S.}~\bibnamefont {Greschner}}, \bibinfo {author} {\bibfnamefont {S.}~\bibnamefont {Mondal}}, \ and\ \bibinfo {author} {\bibfnamefont {T.}~\bibnamefont {Mishra}},\ }\href {\doibase 10.1103/PhysRevA.101.053630} {\bibfield  {journal} {\bibinfo  {journal} {Phys. Rev. A}\ }\textbf {\bibinfo {volume} {101}},\ \bibinfo {pages} {053630} (\bibinfo {year} {2020})}\BibitemShut {NoStop}%
\bibitem [{\citenamefont {Kuno}\ and\ \citenamefont {Hatsugai}(2020)}]{PhysRevResearch.2.042024_interactions}%
  \BibitemOpen
  \bibfield  {author} {\bibinfo {author} {\bibfnamefont {Y.}~\bibnamefont {Kuno}}\ and\ \bibinfo {author} {\bibfnamefont {Y.}~\bibnamefont {Hatsugai}},\ }\href {\doibase 10.1103/PhysRevResearch.2.042024} {\bibfield  {journal} {\bibinfo  {journal} {Phys. Rev. Res.}\ }\textbf {\bibinfo {volume} {2}},\ \bibinfo {pages} {042024} (\bibinfo {year} {2020})}\BibitemShut {NoStop}%
\bibitem [{\citenamefont {Esin}\ \emph {et~al.}(2022)\citenamefont {Esin}, \citenamefont {Kuhlenkamp}, \citenamefont {Refael}, \citenamefont {Berg}, \citenamefont {Rudner},\ and\ \citenamefont {Lindner}}]{esin2022universal_interactions}%
  \BibitemOpen
  \bibfield  {author} {\bibinfo {author} {\bibfnamefont {I.}~\bibnamefont {Esin}}, \bibinfo {author} {\bibfnamefont {C.}~\bibnamefont {Kuhlenkamp}}, \bibinfo {author} {\bibfnamefont {G.}~\bibnamefont {Refael}}, \bibinfo {author} {\bibfnamefont {E.}~\bibnamefont {Berg}}, \bibinfo {author} {\bibfnamefont {M.~S.}\ \bibnamefont {Rudner}}, \ and\ \bibinfo {author} {\bibfnamefont {N.~H.}\ \bibnamefont {Lindner}},\ }\href@noop {} {\bibfield  {journal} {\bibinfo  {journal} {arXiv preprint arXiv:2203.01313}\ } (\bibinfo {year} {2022})}\BibitemShut {NoStop}%
\bibitem [{\citenamefont {Bertok}\ \emph {et~al.}(2022)\citenamefont {Bertok}, \citenamefont {Heidrich-Meisner},\ and\ \citenamefont {Aligia}}]{PhysRevB.106.045141_interactions}%
  \BibitemOpen
  \bibfield  {author} {\bibinfo {author} {\bibfnamefont {E.}~\bibnamefont {Bertok}}, \bibinfo {author} {\bibfnamefont {F.}~\bibnamefont {Heidrich-Meisner}}, \ and\ \bibinfo {author} {\bibfnamefont {A.~A.}\ \bibnamefont {Aligia}},\ }\href {\doibase 10.1103/PhysRevB.106.045141} {\bibfield  {journal} {\bibinfo  {journal} {Phys. Rev. B}\ }\textbf {\bibinfo {volume} {106}},\ \bibinfo {pages} {045141} (\bibinfo {year} {2022})}\BibitemShut {NoStop}%
\bibitem [{\citenamefont {Niu}(1990)}]{PhysRevLett.64.1812_non_adiabatic}%
  \BibitemOpen
  \bibfield  {author} {\bibinfo {author} {\bibfnamefont {Q.}~\bibnamefont {Niu}},\ }\href {\doibase 10.1103/PhysRevLett.64.1812} {\bibfield  {journal} {\bibinfo  {journal} {Phys. Rev. Lett.}\ }\textbf {\bibinfo {volume} {64}},\ \bibinfo {pages} {1812} (\bibinfo {year} {1990})}\BibitemShut {NoStop}%
\bibitem [{\citenamefont {Shih}\ and\ \citenamefont {Niu}(1994)}]{PhysRevB.50.11902_non_adiabatic}%
  \BibitemOpen
  \bibfield  {author} {\bibinfo {author} {\bibfnamefont {W.-K.}\ \bibnamefont {Shih}}\ and\ \bibinfo {author} {\bibfnamefont {Q.}~\bibnamefont {Niu}},\ }\href {\doibase 10.1103/PhysRevB.50.11902} {\bibfield  {journal} {\bibinfo  {journal} {Phys. Rev. B}\ }\textbf {\bibinfo {volume} {50}},\ \bibinfo {pages} {11902} (\bibinfo {year} {1994})}\BibitemShut {NoStop}%
\bibitem [{\citenamefont {Avron}\ and\ \citenamefont {Kons}(1999)}]{avron1999quantum_non_adiabatic}%
  \BibitemOpen
  \bibfield  {author} {\bibinfo {author} {\bibfnamefont {J.~E.}\ \bibnamefont {Avron}}\ and\ \bibinfo {author} {\bibfnamefont {Z.}~\bibnamefont {Kons}},\ }\href@noop {} {\bibfield  {journal} {\bibinfo  {journal} {Journal of Physics A: Mathematical and General}\ }\textbf {\bibinfo {volume} {32}},\ \bibinfo {pages} {6097} (\bibinfo {year} {1999})}\BibitemShut {NoStop}%
\bibitem [{\citenamefont {Privitera}\ \emph {et~al.}(2018)\citenamefont {Privitera}, \citenamefont {Russomanno}, \citenamefont {Citro},\ and\ \citenamefont {Santoro}}]{PhysRevLett.120.106601_non_adiabatic}%
  \BibitemOpen
  \bibfield  {author} {\bibinfo {author} {\bibfnamefont {L.}~\bibnamefont {Privitera}}, \bibinfo {author} {\bibfnamefont {A.}~\bibnamefont {Russomanno}}, \bibinfo {author} {\bibfnamefont {R.}~\bibnamefont {Citro}}, \ and\ \bibinfo {author} {\bibfnamefont {G.~E.}\ \bibnamefont {Santoro}},\ }\href {\doibase 10.1103/PhysRevLett.120.106601} {\bibfield  {journal} {\bibinfo  {journal} {Phys. Rev. Lett.}\ }\textbf {\bibinfo {volume} {120}},\ \bibinfo {pages} {106601} (\bibinfo {year} {2018})}\BibitemShut {NoStop}%
\bibitem [{\citenamefont {Malikis}\ and\ \citenamefont {Cheianov}(2022)}]{10.21468/SciPostPhys.12.6.203_non_adiabatic}%
  \BibitemOpen
  \bibfield  {author} {\bibinfo {author} {\bibfnamefont {S.}~\bibnamefont {Malikis}}\ and\ \bibinfo {author} {\bibfnamefont {V.}~\bibnamefont {Cheianov}},\ }\href {\doibase 10.21468/SciPostPhys.12.6.203} {\bibfield  {journal} {\bibinfo  {journal} {SciPost Phys.}\ }\textbf {\bibinfo {volume} {12}},\ \bibinfo {pages} {203} (\bibinfo {year} {2022})}\BibitemShut {NoStop}%
\bibitem [{\citenamefont {Arceci}\ \emph {et~al.}(2020)\citenamefont {Arceci}, \citenamefont {Kohn}, \citenamefont {Russomanno},\ and\ \citenamefont {Santoro}}]{arceci2020dissipation}%
  \BibitemOpen
  \bibfield  {author} {\bibinfo {author} {\bibfnamefont {L.}~\bibnamefont {Arceci}}, \bibinfo {author} {\bibfnamefont {L.}~\bibnamefont {Kohn}}, \bibinfo {author} {\bibfnamefont {A.}~\bibnamefont {Russomanno}}, \ and\ \bibinfo {author} {\bibfnamefont {G.~E.}\ \bibnamefont {Santoro}},\ }\href@noop {} {\bibfield  {journal} {\bibinfo  {journal} {Journal of Statistical Mechanics: Theory and Experiment}\ }\textbf {\bibinfo {volume} {2020}},\ \bibinfo {pages} {043101} (\bibinfo {year} {2020})}\BibitemShut {NoStop}%
\bibitem [{\citenamefont {Anderson}(1958)}]{PhysRev.109.1492_anderson_loc}%
  \BibitemOpen
  \bibfield  {author} {\bibinfo {author} {\bibfnamefont {P.~W.}\ \bibnamefont {Anderson}},\ }\href {\doibase 10.1103/PhysRev.109.1492} {\bibfield  {journal} {\bibinfo  {journal} {Phys. Rev.}\ }\textbf {\bibinfo {volume} {109}},\ \bibinfo {pages} {1492} (\bibinfo {year} {1958})}\BibitemShut {NoStop}%
\bibitem [{\citenamefont {Lee}\ and\ \citenamefont {Ramakrishnan}(1985)}]{lee1985disordered}%
  \BibitemOpen
  \bibfield  {author} {\bibinfo {author} {\bibfnamefont {P.~A.}\ \bibnamefont {Lee}}\ and\ \bibinfo {author} {\bibfnamefont {T.}~\bibnamefont {Ramakrishnan}},\ }\href@noop {} {\bibfield  {journal} {\bibinfo  {journal} {Reviews of modern physics}\ }\textbf {\bibinfo {volume} {57}},\ \bibinfo {pages} {287} (\bibinfo {year} {1985})}\BibitemShut {NoStop}%
\bibitem [{\citenamefont {Rice}\ and\ \citenamefont {Mele}(1982)}]{PhysRevLett.49.1455_rm_model}%
  \BibitemOpen
  \bibfield  {author} {\bibinfo {author} {\bibfnamefont {M.~J.}\ \bibnamefont {Rice}}\ and\ \bibinfo {author} {\bibfnamefont {E.~J.}\ \bibnamefont {Mele}},\ }\href {\doibase 10.1103/PhysRevLett.49.1455} {\bibfield  {journal} {\bibinfo  {journal} {Phys. Rev. Lett.}\ }\textbf {\bibinfo {volume} {49}},\ \bibinfo {pages} {1455} (\bibinfo {year} {1982})}\BibitemShut {NoStop}%
\bibitem [{\citenamefont {Edwards}\ and\ \citenamefont {Thoules}(1971)}]{edwards1971regularity_anderson_den_states}%
  \BibitemOpen
  \bibfield  {author} {\bibinfo {author} {\bibfnamefont {J.}~\bibnamefont {Edwards}}\ and\ \bibinfo {author} {\bibfnamefont {D.}~\bibnamefont {Thoules}},\ }\href@noop {} {\bibfield  {journal} {\bibinfo  {journal} {Journal of Physics C: Solid State Physics}\ }\textbf {\bibinfo {volume} {4}},\ \bibinfo {pages} {453} (\bibinfo {year} {1971})}\BibitemShut {NoStop}%
\bibitem [{\citenamefont {Lacroix}(1981)}]{lacroix1981density_of_state_anderson}%
  \BibitemOpen
  \bibfield  {author} {\bibinfo {author} {\bibfnamefont {C.}~\bibnamefont {Lacroix}},\ }\href@noop {} {\bibfield  {journal} {\bibinfo  {journal} {Journal of Physics F: Metal Physics}\ }\textbf {\bibinfo {volume} {11}},\ \bibinfo {pages} {2389} (\bibinfo {year} {1981})}\BibitemShut {NoStop}%
\bibitem [{\citenamefont {Biroli}\ \emph {et~al.}(2010)\citenamefont {Biroli}, \citenamefont {Semerjian},\ and\ \citenamefont {Tarzia}}]{biroli2010anderson_lifshitz_tails}%
  \BibitemOpen
  \bibfield  {author} {\bibinfo {author} {\bibfnamefont {G.}~\bibnamefont {Biroli}}, \bibinfo {author} {\bibfnamefont {G.}~\bibnamefont {Semerjian}}, \ and\ \bibinfo {author} {\bibfnamefont {M.}~\bibnamefont {Tarzia}},\ }\href@noop {} {\bibfield  {journal} {\bibinfo  {journal} {Progress of Theoretical Physics Supplement}\ }\textbf {\bibinfo {volume} {184}},\ \bibinfo {pages} {187} (\bibinfo {year} {2010})}\BibitemShut {NoStop}%
\bibitem [{\citenamefont {Johri}\ and\ \citenamefont {Bhatt}(2012)}]{PhysRevLett.109.076402_lifshitz_tails}%
  \BibitemOpen
  \bibfield  {author} {\bibinfo {author} {\bibfnamefont {S.}~\bibnamefont {Johri}}\ and\ \bibinfo {author} {\bibfnamefont {R.~N.}\ \bibnamefont {Bhatt}},\ }\href {\doibase 10.1103/PhysRevLett.109.076402} {\bibfield  {journal} {\bibinfo  {journal} {Phys. Rev. Lett.}\ }\textbf {\bibinfo {volume} {109}},\ \bibinfo {pages} {076402} (\bibinfo {year} {2012})}\BibitemShut {NoStop}%
\bibitem [{\citenamefont {Yaida}(2016)}]{PhysRevB.93.075120_lifshitz_tails}%
  \BibitemOpen
  \bibfield  {author} {\bibinfo {author} {\bibfnamefont {S.}~\bibnamefont {Yaida}},\ }\href {\doibase 10.1103/PhysRevB.93.075120} {\bibfield  {journal} {\bibinfo  {journal} {Phys. Rev. B}\ }\textbf {\bibinfo {volume} {93}},\ \bibinfo {pages} {075120} (\bibinfo {year} {2016})}\BibitemShut {NoStop}%
\bibitem [{\citenamefont {I.M.Lifshitz}(1964)}]{doi:10.1080/00018736400101061}%
  \BibitemOpen
  \bibfield  {author} {\bibinfo {author} {\bibnamefont {I.M.Lifshitz}},\ }\href {\doibase 10.1080/00018736400101061} {\bibfield  {journal} {\bibinfo  {journal} {Advances in Physics}\ }\textbf {\bibinfo {volume} {13}},\ \bibinfo {pages} {483} (\bibinfo {year} {1964})},\ \Eprint {http://arxiv.org/abs/https://doi.org/10.1080/00018736400101061} {https://doi.org/10.1080/00018736400101061} \BibitemShut {NoStop}%
\bibitem [{\citenamefont {Lifshitz}(1965)}]{lifshitz1965energy}%
  \BibitemOpen
  \bibfield  {author} {\bibinfo {author} {\bibfnamefont {I.~M.}\ \bibnamefont {Lifshitz}},\ }\href@noop {} {\bibfield  {journal} {\bibinfo  {journal} {Soviet Physics Uspekhi}\ }\textbf {\bibinfo {volume} {7}},\ \bibinfo {pages} {549} (\bibinfo {year} {1965})}\BibitemShut {NoStop}%
\bibitem [{\citenamefont {Kramer}\ and\ \citenamefont {MacKinnon}(1993)}]{kramer1993localization}%
  \BibitemOpen
  \bibfield  {author} {\bibinfo {author} {\bibfnamefont {B.}~\bibnamefont {Kramer}}\ and\ \bibinfo {author} {\bibfnamefont {A.}~\bibnamefont {MacKinnon}},\ }\href@noop {} {\bibfield  {journal} {\bibinfo  {journal} {Reports on Progress in Physics}\ }\textbf {\bibinfo {volume} {56}},\ \bibinfo {pages} {1469} (\bibinfo {year} {1993})}\BibitemShut {NoStop}%
\bibitem [{\citenamefont {Suominen}(1992)}]{SUOMINEN1992126_lz}%
  \BibitemOpen
  \bibfield  {author} {\bibinfo {author} {\bibfnamefont {K.-A.}\ \bibnamefont {Suominen}},\ }\href {\doibase https://doi.org/10.1016/0030-4018(92)90140-M} {\bibfield  {journal} {\bibinfo  {journal} {Optics Communications}\ }\textbf {\bibinfo {volume} {93}},\ \bibinfo {pages} {126} (\bibinfo {year} {1992})}\BibitemShut {NoStop}%
\bibitem [{\citenamefont {Lehto}\ and\ \citenamefont {Suominen}(2012)}]{PhysRevA.86.033415_lz}%
  \BibitemOpen
  \bibfield  {author} {\bibinfo {author} {\bibfnamefont {J.}~\bibnamefont {Lehto}}\ and\ \bibinfo {author} {\bibfnamefont {K.-A.}\ \bibnamefont {Suominen}},\ }\href {\doibase 10.1103/PhysRevA.86.033415} {\bibfield  {journal} {\bibinfo  {journal} {Phys. Rev. A}\ }\textbf {\bibinfo {volume} {86}},\ \bibinfo {pages} {033415} (\bibinfo {year} {2012})}\BibitemShut {NoStop}%
\bibitem [{\citenamefont {Kam}\ and\ \citenamefont {Chen}(2020)}]{kam2020analytical_lz}%
  \BibitemOpen
  \bibfield  {author} {\bibinfo {author} {\bibfnamefont {C.-F.}\ \bibnamefont {Kam}}\ and\ \bibinfo {author} {\bibfnamefont {Y.}~\bibnamefont {Chen}},\ }\href@noop {} {\bibfield  {journal} {\bibinfo  {journal} {New Journal of Physics}\ }\textbf {\bibinfo {volume} {22}},\ \bibinfo {pages} {023021} (\bibinfo {year} {2020})}\BibitemShut {NoStop}%
\bibitem [{\citenamefont {Altshuler}\ \emph {et~al.}(2010)\citenamefont {Altshuler}, \citenamefont {Krovi},\ and\ \citenamefont {Roland}}]{altshuler2010anderson_local_adiabatic}%
  \BibitemOpen
  \bibfield  {author} {\bibinfo {author} {\bibfnamefont {B.}~\bibnamefont {Altshuler}}, \bibinfo {author} {\bibfnamefont {H.}~\bibnamefont {Krovi}}, \ and\ \bibinfo {author} {\bibfnamefont {J.}~\bibnamefont {Roland}},\ }\href@noop {} {\bibfield  {journal} {\bibinfo  {journal} {Proceedings of the National Academy of Sciences}\ }\textbf {\bibinfo {volume} {107}},\ \bibinfo {pages} {12446} (\bibinfo {year} {2010})}\BibitemShut {NoStop}%
\bibitem [{\citenamefont {Khemani}\ \emph {et~al.}(2015)\citenamefont {Khemani}, \citenamefont {Nandkishore},\ and\ \citenamefont {Sondhi}}]{ved_nand_sond_local_adiabatic}%
  \BibitemOpen
  \bibfield  {author} {\bibinfo {author} {\bibfnamefont {V.}~\bibnamefont {Khemani}}, \bibinfo {author} {\bibfnamefont {R.}~\bibnamefont {Nandkishore}}, \ and\ \bibinfo {author} {\bibfnamefont {S.~L.}\ \bibnamefont {Sondhi}},\ }\href {\doibase 10.1038/nphys3344} {\bibfield  {journal} {\bibinfo  {journal} {Nature Physics}\ }\textbf {\bibinfo {volume} {11}},\ \bibinfo {pages} {560} (\bibinfo {year} {2015})}\BibitemShut {NoStop}%
\bibitem [{\citenamefont {Klitzing}\ \emph {et~al.}(1980)\citenamefont {Klitzing}, \citenamefont {Dorda},\ and\ \citenamefont {Pepper}}]{PhysRevLett.45.494_qhe}%
  \BibitemOpen
  \bibfield  {author} {\bibinfo {author} {\bibfnamefont {K.~v.}\ \bibnamefont {Klitzing}}, \bibinfo {author} {\bibfnamefont {G.}~\bibnamefont {Dorda}}, \ and\ \bibinfo {author} {\bibfnamefont {M.}~\bibnamefont {Pepper}},\ }\href {\doibase 10.1103/PhysRevLett.45.494} {\bibfield  {journal} {\bibinfo  {journal} {Phys. Rev. Lett.}\ }\textbf {\bibinfo {volume} {45}},\ \bibinfo {pages} {494} (\bibinfo {year} {1980})}\BibitemShut {NoStop}%
\bibitem [{\citenamefont {Laughlin}(1981)}]{PhysRevB.23.5632_qhe}%
  \BibitemOpen
  \bibfield  {author} {\bibinfo {author} {\bibfnamefont {R.~B.}\ \bibnamefont {Laughlin}},\ }\href {\doibase 10.1103/PhysRevB.23.5632} {\bibfield  {journal} {\bibinfo  {journal} {Phys. Rev. B}\ }\textbf {\bibinfo {volume} {23}},\ \bibinfo {pages} {5632} (\bibinfo {year} {1981})}\BibitemShut {NoStop}%
\bibitem [{\citenamefont {Thouless}\ \emph {et~al.}(1982)\citenamefont {Thouless}, \citenamefont {Kohmoto}, \citenamefont {Nightingale},\ and\ \citenamefont {den Nijs}}]{PhysRevLett.49.405_qhe}%
  \BibitemOpen
  \bibfield  {author} {\bibinfo {author} {\bibfnamefont {D.~J.}\ \bibnamefont {Thouless}}, \bibinfo {author} {\bibfnamefont {M.}~\bibnamefont {Kohmoto}}, \bibinfo {author} {\bibfnamefont {M.~P.}\ \bibnamefont {Nightingale}}, \ and\ \bibinfo {author} {\bibfnamefont {M.}~\bibnamefont {den Nijs}},\ }\href {\doibase 10.1103/PhysRevLett.49.405} {\bibfield  {journal} {\bibinfo  {journal} {Phys. Rev. Lett.}\ }\textbf {\bibinfo {volume} {49}},\ \bibinfo {pages} {405} (\bibinfo {year} {1982})}\BibitemShut {NoStop}%
\bibitem [{\citenamefont {Sondhi}\ and\ \citenamefont {Yang}(2001)}]{PhysRevB.63.054430_qhe}%
  \BibitemOpen
  \bibfield  {author} {\bibinfo {author} {\bibfnamefont {S.~L.}\ \bibnamefont {Sondhi}}\ and\ \bibinfo {author} {\bibfnamefont {K.}~\bibnamefont {Yang}},\ }\href {\doibase 10.1103/PhysRevB.63.054430} {\bibfield  {journal} {\bibinfo  {journal} {Phys. Rev. B}\ }\textbf {\bibinfo {volume} {63}},\ \bibinfo {pages} {054430} (\bibinfo {year} {2001})}\BibitemShut {NoStop}%
\bibitem [{\citenamefont {Kane}\ \emph {et~al.}(2002)\citenamefont {Kane}, \citenamefont {Mukhopadhyay},\ and\ \citenamefont {Lubensky}}]{PhysRevLett.88.036401_qhe}%
  \BibitemOpen
  \bibfield  {author} {\bibinfo {author} {\bibfnamefont {C.~L.}\ \bibnamefont {Kane}}, \bibinfo {author} {\bibfnamefont {R.}~\bibnamefont {Mukhopadhyay}}, \ and\ \bibinfo {author} {\bibfnamefont {T.~C.}\ \bibnamefont {Lubensky}},\ }\href {\doibase 10.1103/PhysRevLett.88.036401} {\bibfield  {journal} {\bibinfo  {journal} {Phys. Rev. Lett.}\ }\textbf {\bibinfo {volume} {88}},\ \bibinfo {pages} {036401} (\bibinfo {year} {2002})}\BibitemShut {NoStop}%
\bibitem [{\citenamefont {Asada}\ \emph {et~al.}(2002)\citenamefont {Asada}, \citenamefont {Slevin},\ and\ \citenamefont {Ohtsuki}}]{PhysRevLett.89.256601_iqhe_trans}%
  \BibitemOpen
  \bibfield  {author} {\bibinfo {author} {\bibfnamefont {Y.}~\bibnamefont {Asada}}, \bibinfo {author} {\bibfnamefont {K.}~\bibnamefont {Slevin}}, \ and\ \bibinfo {author} {\bibfnamefont {T.}~\bibnamefont {Ohtsuki}},\ }\href {\doibase 10.1103/PhysRevLett.89.256601} {\bibfield  {journal} {\bibinfo  {journal} {Phys. Rev. Lett.}\ }\textbf {\bibinfo {volume} {89}},\ \bibinfo {pages} {256601} (\bibinfo {year} {2002})}\BibitemShut {NoStop}%
\bibitem [{\citenamefont {Slevin}\ and\ \citenamefont {Ohtsuki}(2009)}]{PhysRevB.80.041304_iqhe_trans}%
  \BibitemOpen
  \bibfield  {author} {\bibinfo {author} {\bibfnamefont {K.}~\bibnamefont {Slevin}}\ and\ \bibinfo {author} {\bibfnamefont {T.}~\bibnamefont {Ohtsuki}},\ }\href {\doibase 10.1103/PhysRevB.80.041304} {\bibfield  {journal} {\bibinfo  {journal} {Phys. Rev. B}\ }\textbf {\bibinfo {volume} {80}},\ \bibinfo {pages} {041304} (\bibinfo {year} {2009})}\BibitemShut {NoStop}%
\bibitem [{\citenamefont {Puschmann}\ \emph {et~al.}(2019)\citenamefont {Puschmann}, \citenamefont {Cain}, \citenamefont {Schreiber},\ and\ \citenamefont {Vojta}}]{PhysRevB.99.121301_iqhe_trans}%
  \BibitemOpen
  \bibfield  {author} {\bibinfo {author} {\bibfnamefont {M.}~\bibnamefont {Puschmann}}, \bibinfo {author} {\bibfnamefont {P.}~\bibnamefont {Cain}}, \bibinfo {author} {\bibfnamefont {M.}~\bibnamefont {Schreiber}}, \ and\ \bibinfo {author} {\bibfnamefont {T.}~\bibnamefont {Vojta}},\ }\href {\doibase 10.1103/PhysRevB.99.121301} {\bibfield  {journal} {\bibinfo  {journal} {Phys. Rev. B}\ }\textbf {\bibinfo {volume} {99}},\ \bibinfo {pages} {121301} (\bibinfo {year} {2019})}\BibitemShut {NoStop}%
\bibitem [{\citenamefont {Aubry}\ and\ \citenamefont {Andr{\'e}}(1980)}]{aubry1980analyticity}%
  \BibitemOpen
  \bibfield  {author} {\bibinfo {author} {\bibfnamefont {S.}~\bibnamefont {Aubry}}\ and\ \bibinfo {author} {\bibfnamefont {G.}~\bibnamefont {Andr{\'e}}},\ }\href@noop {} {\bibfield  {journal} {\bibinfo  {journal} {Ann. Israel Phys. Soc}\ }\textbf {\bibinfo {volume} {3}},\ \bibinfo {pages} {18} (\bibinfo {year} {1980})}\BibitemShut {NoStop}%
\bibitem [{\citenamefont {Sambe}(1973)}]{PhysRevA.7.2203_synth_lat}%
  \BibitemOpen
  \bibfield  {author} {\bibinfo {author} {\bibfnamefont {H.}~\bibnamefont {Sambe}},\ }\href {\doibase 10.1103/PhysRevA.7.2203} {\bibfield  {journal} {\bibinfo  {journal} {Phys. Rev. A}\ }\textbf {\bibinfo {volume} {7}},\ \bibinfo {pages} {2203} (\bibinfo {year} {1973})}\BibitemShut {NoStop}%
\bibitem [{\citenamefont {Ho}\ \emph {et~al.}(1983)\citenamefont {Ho}, \citenamefont {Chu},\ and\ \citenamefont {Tietz}}]{HO1983464_synth_lat}%
  \BibitemOpen
  \bibfield  {author} {\bibinfo {author} {\bibfnamefont {T.-S.}\ \bibnamefont {Ho}}, \bibinfo {author} {\bibfnamefont {S.-I.}\ \bibnamefont {Chu}}, \ and\ \bibinfo {author} {\bibfnamefont {J.~V.}\ \bibnamefont {Tietz}},\ }\href {\doibase https://doi.org/10.1016/0009-2614(83)80732-5} {\bibfield  {journal} {\bibinfo  {journal} {Chemical Physics Letters}\ }\textbf {\bibinfo {volume} {96}},\ \bibinfo {pages} {464} (\bibinfo {year} {1983})}\BibitemShut {NoStop}%
\bibitem [{\citenamefont {Verdeny}\ \emph {et~al.}(2016)\citenamefont {Verdeny}, \citenamefont {Puig},\ and\ \citenamefont {Mintert}}]{VerdenyPuigMintert+2016+897+907_synth_lat}%
  \BibitemOpen
  \bibfield  {author} {\bibinfo {author} {\bibfnamefont {A.}~\bibnamefont {Verdeny}}, \bibinfo {author} {\bibfnamefont {J.}~\bibnamefont {Puig}}, \ and\ \bibinfo {author} {\bibfnamefont {F.}~\bibnamefont {Mintert}},\ }\href {\doibase doi:10.1515/zna-2016-0079} {\bibfield  {journal} {\bibinfo  {journal} {Zeitschrift f{\"u}r Naturforschung A}\ }\textbf {\bibinfo {volume} {71}},\ \bibinfo {pages} {897} (\bibinfo {year} {2016})}\BibitemShut {NoStop}%
\bibitem [{\citenamefont {Martin}\ \emph {et~al.}(2017)\citenamefont {Martin}, \citenamefont {Refael},\ and\ \citenamefont {Halperin}}]{PhysRevX.7.041008_synth_latt_qubit}%
  \BibitemOpen
  \bibfield  {author} {\bibinfo {author} {\bibfnamefont {I.}~\bibnamefont {Martin}}, \bibinfo {author} {\bibfnamefont {G.}~\bibnamefont {Refael}}, \ and\ \bibinfo {author} {\bibfnamefont {B.}~\bibnamefont {Halperin}},\ }\href {\doibase 10.1103/PhysRevX.7.041008} {\bibfield  {journal} {\bibinfo  {journal} {Phys. Rev. X}\ }\textbf {\bibinfo {volume} {7}},\ \bibinfo {pages} {041008} (\bibinfo {year} {2017})}\BibitemShut {NoStop}%
\bibitem [{\citenamefont {Peng}\ and\ \citenamefont {Refael}(2018{\natexlab{a}})}]{PhysRevB.97.134303_synth_lat}%
  \BibitemOpen
  \bibfield  {author} {\bibinfo {author} {\bibfnamefont {Y.}~\bibnamefont {Peng}}\ and\ \bibinfo {author} {\bibfnamefont {G.}~\bibnamefont {Refael}},\ }\href {\doibase 10.1103/PhysRevB.97.134303} {\bibfield  {journal} {\bibinfo  {journal} {Phys. Rev. B}\ }\textbf {\bibinfo {volume} {97}},\ \bibinfo {pages} {134303} (\bibinfo {year} {2018}{\natexlab{a}})}\BibitemShut {NoStop}%
\bibitem [{\citenamefont {Peng}\ and\ \citenamefont {Refael}(2018{\natexlab{b}})}]{PhysRevB.98.220509_synth_lat}%
  \BibitemOpen
  \bibfield  {author} {\bibinfo {author} {\bibfnamefont {Y.}~\bibnamefont {Peng}}\ and\ \bibinfo {author} {\bibfnamefont {G.}~\bibnamefont {Refael}},\ }\href {\doibase 10.1103/PhysRevB.98.220509} {\bibfield  {journal} {\bibinfo  {journal} {Phys. Rev. B}\ }\textbf {\bibinfo {volume} {98}},\ \bibinfo {pages} {220509} (\bibinfo {year} {2018}{\natexlab{b}})}\BibitemShut {NoStop}%
\bibitem [{\citenamefont {Crowley}\ \emph {et~al.}(2019)\citenamefont {Crowley}, \citenamefont {Martin},\ and\ \citenamefont {Chandran}}]{PhysRevB.99.064306_synth_lat_qubit}%
  \BibitemOpen
  \bibfield  {author} {\bibinfo {author} {\bibfnamefont {P.~J.~D.}\ \bibnamefont {Crowley}}, \bibinfo {author} {\bibfnamefont {I.}~\bibnamefont {Martin}}, \ and\ \bibinfo {author} {\bibfnamefont {A.}~\bibnamefont {Chandran}},\ }\href {\doibase 10.1103/PhysRevB.99.064306} {\bibfield  {journal} {\bibinfo  {journal} {Phys. Rev. B}\ }\textbf {\bibinfo {volume} {99}},\ \bibinfo {pages} {064306} (\bibinfo {year} {2019})}\BibitemShut {NoStop}%
\bibitem [{\citenamefont {Crowley}\ \emph {et~al.}(2020)\citenamefont {Crowley}, \citenamefont {Martin},\ and\ \citenamefont {Chandran}}]{PhysRevLett.125.100601_synth_lat_qubit}%
  \BibitemOpen
  \bibfield  {author} {\bibinfo {author} {\bibfnamefont {P.~J.~D.}\ \bibnamefont {Crowley}}, \bibinfo {author} {\bibfnamefont {I.}~\bibnamefont {Martin}}, \ and\ \bibinfo {author} {\bibfnamefont {A.}~\bibnamefont {Chandran}},\ }\href {\doibase 10.1103/PhysRevLett.125.100601} {\bibfield  {journal} {\bibinfo  {journal} {Phys. Rev. Lett.}\ }\textbf {\bibinfo {volume} {125}},\ \bibinfo {pages} {100601} (\bibinfo {year} {2020})}\BibitemShut {NoStop}%
\bibitem [{\citenamefont {Long}\ \emph {et~al.}(2021)\citenamefont {Long}, \citenamefont {Crowley},\ and\ \citenamefont {Chandran}}]{PhysRevLett.126.106805_synth_lat}%
  \BibitemOpen
  \bibfield  {author} {\bibinfo {author} {\bibfnamefont {D.~M.}\ \bibnamefont {Long}}, \bibinfo {author} {\bibfnamefont {P.~J.~D.}\ \bibnamefont {Crowley}}, \ and\ \bibinfo {author} {\bibfnamefont {A.}~\bibnamefont {Chandran}},\ }\href {\doibase 10.1103/PhysRevLett.126.106805} {\bibfield  {journal} {\bibinfo  {journal} {Phys. Rev. Lett.}\ }\textbf {\bibinfo {volume} {126}},\ \bibinfo {pages} {106805} (\bibinfo {year} {2021})}\BibitemShut {NoStop}%
\bibitem [{\citenamefont {Nathan}\ \emph {et~al.}(2019)\citenamefont {Nathan}, \citenamefont {Martin},\ and\ \citenamefont {Refael}}]{PhysRevB.99.094311_qubit}%
  \BibitemOpen
  \bibfield  {author} {\bibinfo {author} {\bibfnamefont {F.}~\bibnamefont {Nathan}}, \bibinfo {author} {\bibfnamefont {I.}~\bibnamefont {Martin}}, \ and\ \bibinfo {author} {\bibfnamefont {G.}~\bibnamefont {Refael}},\ }\href {\doibase 10.1103/PhysRevB.99.094311} {\bibfield  {journal} {\bibinfo  {journal} {Phys. Rev. B}\ }\textbf {\bibinfo {volume} {99}},\ \bibinfo {pages} {094311} (\bibinfo {year} {2019})}\BibitemShut {NoStop}%
\bibitem [{\citenamefont {Boyers}\ \emph {et~al.}(2020)\citenamefont {Boyers}, \citenamefont {Crowley}, \citenamefont {Chandran},\ and\ \citenamefont {Sushkov}}]{PhysRevLett.125.160505_synth_lat_qubit}%
  \BibitemOpen
  \bibfield  {author} {\bibinfo {author} {\bibfnamefont {E.}~\bibnamefont {Boyers}}, \bibinfo {author} {\bibfnamefont {P.~J.~D.}\ \bibnamefont {Crowley}}, \bibinfo {author} {\bibfnamefont {A.}~\bibnamefont {Chandran}}, \ and\ \bibinfo {author} {\bibfnamefont {A.~O.}\ \bibnamefont {Sushkov}},\ }\href {\doibase 10.1103/PhysRevLett.125.160505} {\bibfield  {journal} {\bibinfo  {journal} {Phys. Rev. Lett.}\ }\textbf {\bibinfo {volume} {125}},\ \bibinfo {pages} {160505} (\bibinfo {year} {2020})}\BibitemShut {NoStop}%
\bibitem [{\citenamefont {Nathan}\ \emph {et~al.}(2020)\citenamefont {Nathan}, \citenamefont {Refael}, \citenamefont {Rudner},\ and\ \citenamefont {Martin}}]{PhysRevResearch.2.043411_qubit}%
  \BibitemOpen
  \bibfield  {author} {\bibinfo {author} {\bibfnamefont {F.}~\bibnamefont {Nathan}}, \bibinfo {author} {\bibfnamefont {G.}~\bibnamefont {Refael}}, \bibinfo {author} {\bibfnamefont {M.~S.}\ \bibnamefont {Rudner}}, \ and\ \bibinfo {author} {\bibfnamefont {I.}~\bibnamefont {Martin}},\ }\href {\doibase 10.1103/PhysRevResearch.2.043411} {\bibfield  {journal} {\bibinfo  {journal} {Phys. Rev. Res.}\ }\textbf {\bibinfo {volume} {2}},\ \bibinfo {pages} {043411} (\bibinfo {year} {2020})}\BibitemShut {NoStop}%
\bibitem [{\citenamefont {Long}\ \emph {et~al.}(2022{\natexlab{a}})\citenamefont {Long}, \citenamefont {Crowley}, \citenamefont {Koll\'ar},\ and\ \citenamefont {Chandran}}]{PhysRevLett.128.183602_qubit}%
  \BibitemOpen
  \bibfield  {author} {\bibinfo {author} {\bibfnamefont {D.~M.}\ \bibnamefont {Long}}, \bibinfo {author} {\bibfnamefont {P.~J.~D.}\ \bibnamefont {Crowley}}, \bibinfo {author} {\bibfnamefont {A.~J.}\ \bibnamefont {Koll\'ar}}, \ and\ \bibinfo {author} {\bibfnamefont {A.}~\bibnamefont {Chandran}},\ }\href {\doibase 10.1103/PhysRevLett.128.183602} {\bibfield  {journal} {\bibinfo  {journal} {Phys. Rev. Lett.}\ }\textbf {\bibinfo {volume} {128}},\ \bibinfo {pages} {183602} (\bibinfo {year} {2022}{\natexlab{a}})}\BibitemShut {NoStop}%
\bibitem [{\citenamefont {Vuina}\ \emph {et~al.}(2023)\citenamefont {Vuina}, \citenamefont {Long}, \citenamefont {Crowley},\ and\ \citenamefont {Chandran}}]{PhysRevB.108.134303_qubit}%
  \BibitemOpen
  \bibfield  {author} {\bibinfo {author} {\bibfnamefont {D.}~\bibnamefont {Vuina}}, \bibinfo {author} {\bibfnamefont {D.~M.}\ \bibnamefont {Long}}, \bibinfo {author} {\bibfnamefont {P.~J.~D.}\ \bibnamefont {Crowley}}, \ and\ \bibinfo {author} {\bibfnamefont {A.}~\bibnamefont {Chandran}},\ }\href {\doibase 10.1103/PhysRevB.108.134303} {\bibfield  {journal} {\bibinfo  {journal} {Phys. Rev. B}\ }\textbf {\bibinfo {volume} {108}},\ \bibinfo {pages} {134303} (\bibinfo {year} {2023})}\BibitemShut {NoStop}%
\bibitem [{\citenamefont {Weinberg}\ and\ \citenamefont {Bukov}(2017)}]{10.21468/SciPostPhys.2.1.003_quspin}%
  \BibitemOpen
  \bibfield  {author} {\bibinfo {author} {\bibfnamefont {P.}~\bibnamefont {Weinberg}}\ and\ \bibinfo {author} {\bibfnamefont {M.}~\bibnamefont {Bukov}},\ }\href {\doibase 10.21468/SciPostPhys.2.1.003} {\bibfield  {journal} {\bibinfo  {journal} {SciPost Phys.}\ }\textbf {\bibinfo {volume} {2}},\ \bibinfo {pages} {003} (\bibinfo {year} {2017})}\BibitemShut {NoStop}%
\bibitem [{\citenamefont {Weinberg}\ and\ \citenamefont {Bukov}(2019)}]{10.21468/SciPostPhys.7.2.020_quspin}%
  \BibitemOpen
  \bibfield  {author} {\bibinfo {author} {\bibfnamefont {P.}~\bibnamefont {Weinberg}}\ and\ \bibinfo {author} {\bibfnamefont {M.}~\bibnamefont {Bukov}},\ }\href {\doibase 10.21468/SciPostPhys.7.2.020} {\bibfield  {journal} {\bibinfo  {journal} {SciPost Phys.}\ }\textbf {\bibinfo {volume} {7}},\ \bibinfo {pages} {020} (\bibinfo {year} {2019})}\BibitemShut {NoStop}%
\bibitem [{\citenamefont {Landau}\ and\ \citenamefont {Sowjetunion}(1932)}]{landau19322}%
  \BibitemOpen
  \bibfield  {author} {\bibinfo {author} {\bibfnamefont {L.}~\bibnamefont {Landau}}\ and\ \bibinfo {author} {\bibfnamefont {Z.}~\bibnamefont {Sowjetunion}},\ }\href@noop {} {\bibfield  {journal} {\bibinfo  {journal} {Proc. R. Soc. London A}\ }\textbf {\bibinfo {volume} {137}},\ \bibinfo {pages} {696} (\bibinfo {year} {1932})}\BibitemShut {NoStop}%
\bibitem [{\citenamefont {Majorana}(1932)}]{majorana1932atomi}%
  \BibitemOpen
  \bibfield  {author} {\bibinfo {author} {\bibfnamefont {E.}~\bibnamefont {Majorana}},\ }\href@noop {} {\bibfield  {journal} {\bibinfo  {journal} {Il Nuovo Cimento (1924-1942)}\ }\textbf {\bibinfo {volume} {9}},\ \bibinfo {pages} {43} (\bibinfo {year} {1932})}\BibitemShut {NoStop}%
\bibitem [{\citenamefont {St{\"u}ckelberg}(1932)}]{stuckelberg1932theorie}%
  \BibitemOpen
  \bibfield  {author} {\bibinfo {author} {\bibfnamefont {E.}~\bibnamefont {St{\"u}ckelberg}},\ }\href@noop {} {\bibfield  {journal} {\bibinfo  {journal} {Helv. Phys. Acta}\ }\textbf {\bibinfo {volume} {5}},\ \bibinfo {pages} {369} (\bibinfo {year} {1932})}\BibitemShut {NoStop}%
\bibitem [{\citenamefont {Bachmann}\ \emph {et~al.}(2017)\citenamefont {Bachmann}, \citenamefont {De~Roeck},\ and\ \citenamefont {Fraas}}]{PhysRevLett.119.060201_ad_theorem}%
  \BibitemOpen
  \bibfield  {author} {\bibinfo {author} {\bibfnamefont {S.}~\bibnamefont {Bachmann}}, \bibinfo {author} {\bibfnamefont {W.}~\bibnamefont {De~Roeck}}, \ and\ \bibinfo {author} {\bibfnamefont {M.}~\bibnamefont {Fraas}},\ }\href {\doibase 10.1103/PhysRevLett.119.060201} {\bibfield  {journal} {\bibinfo  {journal} {Phys. Rev. Lett.}\ }\textbf {\bibinfo {volume} {119}},\ \bibinfo {pages} {060201} (\bibinfo {year} {2017})}\BibitemShut {NoStop}%
\bibitem [{\citenamefont {Bachmann}\ \emph {et~al.}(2020)\citenamefont {Bachmann}, \citenamefont {De~Roeck},\ and\ \citenamefont {Fraas}}]{bachmann2020adiabatic_ad_theorem}%
  \BibitemOpen
  \bibfield  {author} {\bibinfo {author} {\bibfnamefont {S.}~\bibnamefont {Bachmann}}, \bibinfo {author} {\bibfnamefont {W.}~\bibnamefont {De~Roeck}}, \ and\ \bibinfo {author} {\bibfnamefont {M.}~\bibnamefont {Fraas}},\ }\href@noop {} {\bibfield  {journal} {\bibinfo  {journal} {Analytic Trends in Mathematical Physics}\ }\textbf {\bibinfo {volume} {741}},\ \bibinfo {pages} {43} (\bibinfo {year} {2020})}\BibitemShut {NoStop}%
\bibitem [{\citenamefont {Hughes}\ \emph {et~al.}(1994)\citenamefont {Hughes}, \citenamefont {Nicholls}, \citenamefont {Frost}, \citenamefont {Linfield}, \citenamefont {Pepper}, \citenamefont {Ford}, \citenamefont {Ritchie}, \citenamefont {Jones}, \citenamefont {Kogan},\ and\ \citenamefont {Kaveh}}]{hughes1994magnetic_iqhe_trans}%
  \BibitemOpen
  \bibfield  {author} {\bibinfo {author} {\bibfnamefont {R.}~\bibnamefont {Hughes}}, \bibinfo {author} {\bibfnamefont {J.}~\bibnamefont {Nicholls}}, \bibinfo {author} {\bibfnamefont {J.}~\bibnamefont {Frost}}, \bibinfo {author} {\bibfnamefont {E.}~\bibnamefont {Linfield}}, \bibinfo {author} {\bibfnamefont {M.}~\bibnamefont {Pepper}}, \bibinfo {author} {\bibfnamefont {C.}~\bibnamefont {Ford}}, \bibinfo {author} {\bibfnamefont {D.}~\bibnamefont {Ritchie}}, \bibinfo {author} {\bibfnamefont {G.}~\bibnamefont {Jones}}, \bibinfo {author} {\bibfnamefont {E.}~\bibnamefont {Kogan}}, \ and\ \bibinfo {author} {\bibfnamefont {M.}~\bibnamefont {Kaveh}},\ }\href@noop {} {\bibfield  {journal} {\bibinfo  {journal} {Journal of Physics: Condensed Matter}\ }\textbf {\bibinfo {volume} {6}},\ \bibinfo {pages} {4763} (\bibinfo {year} {1994})}\BibitemShut {NoStop}%
\bibitem [{\citenamefont {Yang}\ and\ \citenamefont {Bhatt}(1996)}]{PhysRevLett.76.1316_iqhe_trans}%
  \BibitemOpen
  \bibfield  {author} {\bibinfo {author} {\bibfnamefont {K.}~\bibnamefont {Yang}}\ and\ \bibinfo {author} {\bibfnamefont {R.~N.}\ \bibnamefont {Bhatt}},\ }\href {\doibase 10.1103/PhysRevLett.76.1316} {\bibfield  {journal} {\bibinfo  {journal} {Phys. Rev. Lett.}\ }\textbf {\bibinfo {volume} {76}},\ \bibinfo {pages} {1316} (\bibinfo {year} {1996})}\BibitemShut {NoStop}%
\bibitem [{\citenamefont {Chalker}\ and\ \citenamefont {Coddington}(1988)}]{chalker1988percolation_iqhe_trans}%
  \BibitemOpen
  \bibfield  {author} {\bibinfo {author} {\bibfnamefont {J.}~\bibnamefont {Chalker}}\ and\ \bibinfo {author} {\bibfnamefont {P.}~\bibnamefont {Coddington}},\ }\href@noop {} {\bibfield  {journal} {\bibinfo  {journal} {Journal of Physics C: Solid State Physics}\ }\textbf {\bibinfo {volume} {21}},\ \bibinfo {pages} {2665} (\bibinfo {year} {1988})}\BibitemShut {NoStop}%
\bibitem [{\citenamefont {Long}\ \emph {et~al.}(2022{\natexlab{b}})\citenamefont {Long}, \citenamefont {Crowley},\ and\ \citenamefont {Chandran}}]{PhysRevB.105.144204_MBL_qp}%
  \BibitemOpen
  \bibfield  {author} {\bibinfo {author} {\bibfnamefont {D.~M.}\ \bibnamefont {Long}}, \bibinfo {author} {\bibfnamefont {P.~J.~D.}\ \bibnamefont {Crowley}}, \ and\ \bibinfo {author} {\bibfnamefont {A.}~\bibnamefont {Chandran}},\ }\href {\doibase 10.1103/PhysRevB.105.144204} {\bibfield  {journal} {\bibinfo  {journal} {Phys. Rev. B}\ }\textbf {\bibinfo {volume} {105}},\ \bibinfo {pages} {144204} (\bibinfo {year} {2022}{\natexlab{b}})}\BibitemShut {NoStop}%
\end{thebibliography}%

\appendix

\section{Excitation rates}

The excitation rate between the bands can be calculated from the transition rate between individual states belonging to the two bands~\eqref{eq:exc_rate}. The two scattering mechanisms discussed in the main text---adiabatic and non-adiabatic---yield distinct scalings of the excitation rate and hence the charge pumping lifetime. 

In both cases the excitation rate may be expressed as
\begin{equation}\label{eq:exc_rate_app}
\begin{split}
    \Gamma =\omega\int \mathrm{d}r_- \, \mathrm{d}r_+ \, \mathrm{d}E_- \, \mathrm{d}E_+ \, n(E_-) \, n(E_+)  P_{\mathrm{exc}},
\end{split}
\end{equation}
where $P_{\mathrm{exc}}$ is the excitation probability in a single period between two states with localisation centers $r_{\pm}$ and energy $E_{\pm}$, and $n(E_{\pm})$ is the density of states. The dominant contribution to the scattering rate for both mechanisms comes from the states near the bottom of the bands, but $P_{\mathrm{exc}}$ will have a different form in each case. For our class of models~\eqref{eq:model_eq}, with extensive perturbations, the scattering rate~\eqref{eq:exc_rate_app} will also be extensive, as $\mathcal{O}(L)$ states can scatter. 

In Sec.~\ref{sec:non-adiabatic} we evaluate Eq.~\eqref{eq:exc_rate_app} for the non-adiabatic scattering mechanism at small $\omega$, while the adiabatic scattering mechanism is treated in Sec.~\ref{sec:adiabatic}. 

\subsection{Non-adiabatic mechanism}\label{sec:non-adiabatic}
The transitions between states of two bands in the non-adiabatic mechanism are modelled by a parabolic level crossing
\begin{equation}\label{eq:effective_two_state_problem_app}
\begin{split}
    H^{\mathrm{eff}}(t-t^*) = \left(\epsilon_0 + \kappa\omega^2 (t-t^*)^2\right)\sigma_z + V\sigma_x + \\ \mathcal{O}\left( \omega^3 (t-t^*)^3\right).
\end{split}
\end{equation}
We take the matrix element $V$ to be exponentially small in the distance between the localisation centers 
\begin{equation}\label{eq:matrix_ele_app}
    V=\mathcal{O}\left(Je^{ -|r_+ - r_-|/\max\{\xi_{\mathrm{loc}}(E_+), \xi_{\mathrm{loc}}(E_-)\}}\right),
\end{equation}
where $\xi_{\pm}(E_{\pm})$ are the localisation lengths of instantaneous eigenstates in the upper and lower bands. 

The parabolic crossing problem is not fully solvable, so must resort to approximations to estimate the transition probability $P_{\mathrm{exc}}$. We will consider $\epsilon_0 \gg V$. 

Indeed, for $W<W_c$ and not too close to $W_c$ this condition holds. The band edges are given by Eq.~\eqref{eq:band_edge}, so the smallest instantaneous energy gap is finite $\epsilon_0 \geq W_c-W$. Meanwhile, the matrix elements are exponentially small in distance between localisation centers. On the other hand, for $W>W_c$, the two bands of states overlap, and the adiabatic mechanism (Sec.~\eqref{sec:adiabatic}) takes over.

In the limit $\epsilon_0 \gg V$, Ref.~\cite{SUOMINEN1992126_lz} shows that the transition probability is given by
\begin{equation}\label{eq:trans_prob_non_ad_app}
    P_{\mathrm{exc}}(\epsilon_0, \omega, V) = \frac{\pi V^2}{2 \omega \sqrt{\kappa \epsilon_0}} \exp(-\frac{8}{3}\frac{\epsilon_0^{3/2}}{\omega \sqrt{\kappa}}).
\end{equation}
Because the excitation probability is super-exponentially small in $\epsilon_0$, the dominant contribution to the excitation rate~\eqref{eq:exc_rate_app} comes from the band edges. 

The matrix elements~\eqref{eq:matrix_ele_app} are simplified for states near the band edges. Since the localisation length of states is a smooth function of energy (which does not diverge at the band edge)~\cite{PhysRevLett.109.076402_lifshitz_tails}, we may approximate it as $\xi(E_{\pm})=\xi$. Equation~\eqref{eq:matrix_ele_app} becomes $V=\mathcal{O}\left(Je^{ -|r_+ - r_-|/\xi}\right)$. Note that the states at the band edges are confined to rare regions of contiguous sites with low disorder. The spatial extent of the rare regions is much smaller than their typical spatial separation. This validates considering rare region states as genuinely localised states.

Evaluating the integrals over the distances in the bands, to leading order in the $\omega \to 0$ and $V \ll \epsilon_0$ limits, the scattering rate~\eqref{eq:exc_rate_app} becomes 
\begin{equation}
\begin{split}
    \Gamma = \xi L \frac{J^{2}\pi}{4 \sqrt{\kappa}} \int \, \mathrm{d}E_- \, \mathrm{d}E_+ \, n(E_-) \, n(E_+) \sqrt{\frac{2}{E_+-E_-}} \\
    \times \exp(-\frac{8|(E_+-E_-)/2|^{3/2}}{3\omega\sqrt{\kappa}}).
\end{split}
\end{equation}

We calculate the integrals over the energies in the two bands by restricting them to the band edges $E_{\pm}=\pm(\delta + e_{\pm})$, where $|e_{\pm}| \ll \delta$. Here, $\delta$ is half of the instantaneous band gap at its smallest value during a cycle. The density of states at the edges of the two bands is given by the Lifshitz tail form $n(E_{\pm}) = Ae^{-R|\pm \delta - E_{\pm}|^{-1/2}}$, where $A$ is a normalisation constant with units of inverse energy and inverse length. Inputting the appropriate integration windows the scattering rate becomes
\begin{equation}
\begin{split}
    \Gamma = \xi L A^2 \frac{J^{2}\pi}{4 \sqrt{\kappa}}\int \, \mathrm{d}e_- \, \mathrm{d}e_+ \, e^{-Re_+^{-1/2}} \, e^{-Re_-^{-1/2}} \, \\
    \sqrt{\frac{2}{2\delta + e_+ + e_-}}\exp(-\frac{8|\delta +( e_+ + e_-)/2|^{3/2}}{3\omega\sqrt{\kappa}}).
\end{split}
\end{equation}
The energy integrals above can be done by saddle point approximation. Evaluating these integrals to first order approximation in $e_{\pm} \ll \delta$, we obtain
\begin{equation}\label{eq:exc_rate3}
\begin{split}
    \Gamma = \xi L A^2 \frac{J^{2}\pi R^4}{4 \sqrt{\kappa \delta}}  \,  \exp(-\frac{8\delta^{3/2}}{3\omega\sqrt{\kappa}} -(4R)^{2/3}\left(\frac{\delta}{\kappa}\right)^{1/6}\omega^{-1/3} ).
\end{split}
\end{equation}
Note that the saddle points $e_{\pm}=R^2 \left(\frac{\omega^2\kappa}{16R^4\delta}\right)^{1/3}$ indeed go to $0$ as $\omega \to 0$. This is consistent with the claim that the scattering rate is dominated by the band edge states in the asymptotic limit $\omega \to 0$. The second term in the exponent is subdominant in the limit $\omega \to 0$, so we drop it in the scaling analysis in Eq.~\eqref{eq:omega_loc_length}.

We estimate the parameters $\delta$ and $\kappa$ in the approximation to the scattering rate in Eq.~\eqref{eq:exc_rate3}. Parameters $\delta$ and $\kappa$ correspond to the minimal eigenenergy separation of the states in the opposite bands during a cycle and the curvature of the eigenenergy at the band edges, respectively. To calculate these we expand the eigenenergy of the bands at their extrema during the cycle $t=0$
\begin{equation}\label{eq:dis_eigener_exp}
\begin{split}
    E^{\mathrm{dis}}_{\pm}(k=\pi/2,t) = \pm \Bigg[ \sqrt{2}\sqrt{\delta_0^2+J^2} \\ + \frac{\Delta_0^2-2 \delta_0^2}{2\sqrt{2}\sqrt{\delta_0^2+J^2}} \omega^2 t^2 + \mathcal{O}\left( \left(\omega t \right)^3 \right) \Bigg] \mp W/2.
\end{split}
\end{equation}
Here we assume that the edges of both bands are broadened by disorder according to Eq.~\eqref{eq:band_edge}. In this model, we can even make a prediction of the critical disorder as being the value of $W$ at which $\delta=0$. We have $\delta = \sqrt{2}\sqrt{\delta_0^2+J^2} - W/2=(W_c-W)/2$, yielding the critical disorder $W_c \approx 3.16J$ for $\delta_0=0.5 J$, as in the main text. This critical disorder strength is in good agreement with numerics. 

The coefficient of $\omega^2 t^2$ in Eq.~\eqref{eq:dis_eigener_exp} is the curvature
\begin{equation}
    \kappa=\frac{\Delta_0^2-2 \delta_0^2}{2\sqrt{2}\sqrt{\delta_0^2+J^2}} \approx 0.55 J, 
\end{equation}
for $\delta_0=0.5 J$ and $\Delta_0=1.5 J$, as in the main text. The full scattering rate is
\begin{equation}\label{eq:full_scattering_rate_appendix}
    \Gamma = \xi L A^2\frac{\sqrt{2}J^{2}\pi R^4}{4 \sqrt{\kappa (W-W_c)}}  \,  \exp(-\frac{kJ}{\omega}),
\end{equation}
where $k \approx 1.27\left(\frac{W-W_c}{J}\right)^{3/2}$. This estimate of $k$ is close to the value used for the scaling collapse in Fig.~\ref{fig:scaling_fig}, but we were unable to confirm the scaling $k \propto (W-W_c)^{3/2}$.

\subsection{Adiabatic mechanism}\label{sec:adiabatic}

The adiabatic mechanism controls the scattering rate~\eqref{eq:exc_rate_app} as the band gap closes ($W \geq W_c$) in the adiabatic limit. Near the transition $(W \to W_c)$ the states that participate in scattering are at the band edges. 

We estimate the probability of excitation per period $P_{\mathrm{exc}}$ due to the adiabatic mechanism. There is a finite time window in which the two bands coalesce in energy. Each state of the lower band forms many level crossings with states of the upper band in this window. Predicting the population of the upper band at the end on the window is a complicated biased tree problem. A simplification to this problem is estimating an average excitation probability between the bands per period.

The average probability of excitation into the upper band is finite, $P_{\mathrm{exc}}=\mathcal{O}(1)$. We model each crossing as involving only two levels with Landau-Zener crossing~\cite{landau19322}, 
\begin{equation}\label{eq:lz}
    H(t) = V \sigma_x + \kappa' \omega t \sigma_z.
\end{equation}
The gaps at these crossings are set by the matrix elements $V$ connecting the instantaneous states. If the gap between the states is much larger than the drive frequency $V \gg \omega$ the probability of excitation to the upper band is negligible. The adiabatic theorem~\cite{landau19322, majorana1932atomi,stuckelberg1932theorie,PhysRevLett.119.060201_ad_theorem,bachmann2020adiabatic_ad_theorem} states the weight will remain in the lower branch at every crossing ending up in the same band. In the opposite limit $V \ll \omega$ the probability of excitation to the upper band is also negligible since the diabatic transitions lead to the weight remaining in the state belonging to the lower band until it leaves the overlap region. However, a crossing with a gap $V=\mathcal{O}(\omega)$ leads to splitting of weight between the two branches. A single excitation like this leads to the finite probability of excitation to the upper band. In the order of limits $L \to \infty$, $\omega \to 0$ there are always gaps as small as the drive frequency, so that $V = \mathcal{O}(\omega)$ crossings always occur no matter how small $\omega$ is~\cite{altshuler2010anderson_local_adiabatic,ved_nand_sond_local_adiabatic}. 

We compute the scattering rate~\eqref{eq:exc_rate_app} by counting the number of states participating in level crossings with $V=\mathcal{O}(\omega)$. We define the dimensionless number $\zeta=\mathcal{O}(1)$, such that the crossings with matrix elements in the range $V \in \left[ \omega e^{-\zeta}, \omega e^{\zeta} \right]$ are considered. The spatial separation between such states is $r \in  r_0 + \xi \left[  -\zeta, \zeta \right]$ with $r_0 = \xi \log \left(J/\omega \right)$. For every state in the lower band the number of such states in the upper band, within an energy window $\mathrm{d}E$, is  
\begin{equation}
     \int_{r_0-\zeta}^{r_0+\zeta} \, \mathrm{d}r_+ \, \mathrm{d}E \, n_+(E) = 2 \xi \zeta n_+(E) \mathrm{d}E,
\end{equation}
where $n_+$ is the density of states of the upper band at energy $E$. We normalise the density of states such that its integral over distance and energy is 1. That is, distance and energy are measured in the units of the number of unit cells and bandwidth, respectively.

The number of states in the lower band in the energy window $\mathrm{d}E$ at energy $E$ is 
\begin{equation}
     \int \, \mathrm{d}r_- \, \mathrm{d}E \, n_-(E) = L n_-(E)\mathrm{d}E.
\end{equation}

Piecing the scattering rate~\eqref{eq:exc_rate_app} together gives
\begin{equation}\label{eq:exc_rate2}
\begin{split}
    \Gamma =2 \xi \zeta L \omega \int \, \mathrm{d}E_+ \, \mathrm{d}E_- \, n_+(E_+) n_-(E_-).
\end{split}
\end{equation}
Inserting the density of states near the band edges and integrating over the overlapping energy region  $E_{\pm} \in \left[ -\frac{W-W_c}{2}, \frac{W-W_c}{2} \right]$ by saddle point approximation we obtain
\begin{equation}\label{eq:exc_rate3A}
\begin{split}
    \Gamma =2 \xi \zeta L \omega R^4\exp \left[ -2R\left(W-W_c\right)^{-1/2} \right].
\end{split}
\end{equation}
Note that the rate~\eqref{eq:exc_rate3A} is only valid as we approach the transition from above the critical disorder value.

\section{Adiabatic limit transition}
We predict that the adiabatic transition at $W=W_c$ occurs when the gap between the instantaneous bands closes at least once per cycle of the drive. At the transition the excitations between the instantaneous eigenstates at the band edges cause the breakdown of charge pumping. The excitation rate per period due to this process gives the pumping decay timescale in the thermodynamic limit~\eqref{eq:omega_loc_length}. At finite system sizes the decay in charge pumping is cut off as the charges reach the boundary of the system. Therefore, the pump rate $\bar{Q}$ assumes a finite value across the $W=W_c$ transition at finite system sizes. This allows us to perform finite-size scaling analysis on the disorder averaged parameter $\bar{Q}$ across the transition in the adiabatic limit. 

The finite-size scaling is done by comparing the length of the system $L$ to the length scale over which the charge is transferred 
\begin{equation}
    \xi_W = \tau_Q/T \sim \exp\left[ 2R\left( W-W_c \right)^{-1/2}\right]
\end{equation} 
measured in the number of unit cells. Here, $\tau_Q$ is the timescale associated with the decay in pumping. This timescale~\eqref{eq:loc_time_W} is the inverse of the scattering rate~\eqref{eq:exc_rate3A} per state. 

\begin{figure}
    \centering
    \includegraphics[width=\linewidth]{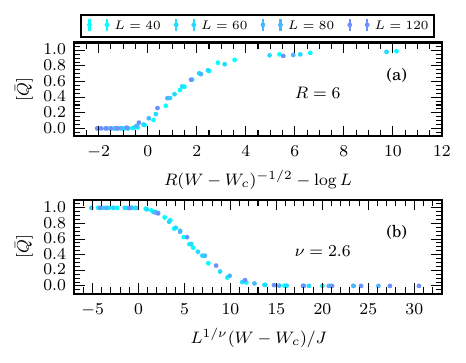}
    \caption{Finite size scaling analysis~\eqref{eq:scaling_collapse_W} of $\bar{Q}$ using data in Fig.~\ref{fig:rare_r_physics} (a). (a) Rescaling of the $W$ axis according to the adiabatic scattering mechanism prediction for the length scale over which charge is transported $\xi_W$ in Eq.~\eqref{eq:loc_length_W} with $R=6$. (b) Rescaling the $W$ axis according to the IQHE prediction for the divergence of the steady state localisation length $\xi$ in Eq.~\eqref{eq:IQHE_loc} with $\nu=2.6$. The critical point for both rescalings is $W_c \approx 3.16 J$. Numerical data is consistent with both scaling forms.}
    \label{fig:scaling_W}
\end{figure}

In the vicinity the transition, the long-time averaged charge pumped per period becomes a function of a single variable
\begin{equation}\label{eq:scaling_collapse_W}
    \bar{Q} \sim \Tilde{Q}\left( \xi_W/L \right).
\end{equation}
To check this scaling relationship, we use the data from Fig.~\ref{fig:rare_r_physics} (a) and plot it in Fig.~\ref{fig:scaling_W} (a) with $\log(\xi_W/L)$ on the horizontal axis. The scaling relationship~\eqref{eq:scaling_collapse_W} is consistent with data from Fig.~\ref{fig:rare_r_physics} (a). The parameter $R$, found in the scaling of Lifshitz tail states with energy~\eqref{eq:den_states_tails} has not been found numerically. We leave it as a free parameter. Note that the data with $[\bar{Q}]=1$ for $W<W_c$ from Fig.~\ref{fig:rare_r_physics} (a) is not included in Fig.~\ref{fig:scaling_W} (a) since the length scale $\xi_W$ is not well defined there. 

Our understanding of the Thouless pumping transition in the adiabatic limit is notably different from~\cite{PhysRevLett.123.266601_disorder}. This work studied the disorder induced transition in the Thouless pump as a localisation-delocalisation transition of the steady states of the system. Computing the localisation length of the steady states across the transition, they obtain a good scaling collapse with the localisation length scaling as
\begin{equation}\label{eq:IQHE_loc}
    \xi \sim |(W-W_c)/J|^{-1/\nu},
\end{equation}
where $\nu \approx 2$. This relationship was derived assuming the transition is in the universality class of the integer quantum Hall effect (IQHE) transition~\cite{hughes1994magnetic_iqhe_trans, PhysRevB.80.041304_iqhe_trans, PhysRevB.99.121301_iqhe_trans, PhysRevLett.76.1316_iqhe_trans, PhysRevLett.89.256601_iqhe_trans, chalker1988percolation_iqhe_trans}. Figure~\ref{fig:scaling_W} (b) tests the scaling~\eqref{eq:scaling_collapse_W} using~\eqref{eq:IQHE_loc} instead of $\xi_W$ with $\nu=2.6$~\cite{PhysRevB.80.041304_iqhe_trans}. This collapse is also consistent with the data from Fig.~\ref{fig:rare_r_physics} (a). 

At these system sizes, finite size scaling analysis is unable to distinguish between the different proposed universality classes of the Thouless pump transition in the adiabatic limit. At larger system sizes the Lifshitz state physics should be more prominent. Accessing the adiabatic limit at very large system sizes would be helpful in distinguishing between these universality classes.

\section{Connections to other physical systems}

The similarity of synthetic lattice constructions for the disordered Thouless pump, the two-tone driven qubit and the disordered Chern insulator model in an electric field allow parallels to be drawn between the dynamical phenomena in these systems~\cite{PhysRevB.99.094311_qubit, PhysRevB.108.134303_qubit, PhysRevLett.128.183602_qubit, PhysRevResearch.2.043411_qubit, PhysRevB.99.064306_synth_lat_qubit, PhysRevX.7.041008_synth_latt_qubit, PhysRevLett.125.100601_synth_lat_qubit,thouless1998topological_exponentially_small_corr,PhysRevB.35.2188_exponentially_small_corr,PhysRevResearch.2.042035_disorder,PhysRevLett.124.086602_disorder,PhysRevLett.123.266601_disorder,PhysRevA.106.L051301_disorder,PhysRevB.100.184304_disorder,PhysRevA.103.043310_disorder,PhysRevLett.125.160505_synth_lat_qubit,PhysRevA.101.052323_disorder}. 

The synthetic lattice construction of the disordered Thouless pump is made by Fourier transforming the Schrodinger equation 
\begin{equation}\label{eq:Sch_eq}
    i \partial_t | \psi(t) \rangle = H(t) | \psi(t) \rangle
\end{equation}
for the model in Eq.~\eqref{eq:model_eq}. We look for a complete set of solutions to Schrodinger equation~\eqref{eq:Sch_eq} of the form $|\psi_{\alpha}(t)\rangle = e^{-i \epsilon_{\alpha} t} |\phi_{\alpha}(t) \rangle $, where $|\phi_{\alpha}(t+T)\rangle = |\phi_{\alpha} (t) \rangle$ is periodic and $\alpha$ labels the basis of the Hilbert space of the chain. 
Rewriting $H(t) = H(\theta_t)$ with $\theta_t = \omega t  + \theta_0$ defined modulo $2 \pi$, we decompose $H$ into its Fourier components 
\begin{equation}
    H(\theta) = \sum_{m \in \mathbb{Z}} H_m e^{im \theta}.
\end{equation}
Expanding $|\phi_{\alpha}(\theta_t) \rangle = \sum_{m \in \mathbb{Z}} |\phi_{\alpha m}(\theta_0) \rangle e^{im \theta_t}$, the Schrodinger equation transforms to 
\begin{equation}\label{eq:freq_latt_eq}
\begin{split}
    \epsilon_{\alpha} |\phi_{\alpha n}(\theta_0) \rangle =\sum_{m \in \mathbb{Z}} \left(H_{n-m} e^{i(n-m)\theta_0} + n \omega \delta_{mn} \right) \\
    |\phi_{\alpha m}(\theta_0) \rangle.
\end{split}
\end{equation}

Defining an auxiliary Hilbert space spanned by $|n \rangle$ casts Eq.~\eqref{eq:freq_latt_eq} as an eigenvalue equation for $K(\theta_0)$, with
\begin{equation}\label{eq:frequency_lattice}
    \begin{split}
        &|\Tilde{\phi}_{\alpha}(\theta_0) \rangle = \sum_{n \in \mathbb{Z}} |\phi_{\alpha n}(\theta_0) \rangle \otimes |n \rangle, 
        \\
        &K(\theta_0) = \sum_{m, n \, \in \, \mathbb{Z}} \left(H_{n-m} e^{i(n-m)\theta_0} + n \omega \delta_{mn} \right) \otimes |n \rangle \langle m|.
    \end{split}
\end{equation}
The eigenvalue equation~\eqref{eq:freq_latt_eq} is of the form of a tight-binding model on a 2D lattice with sites labelled by $n$ and the original spatial dimension of the chain. The extra dimension in the effective lattice model $K$ has a natural interpretation as the number of drive quanta absorbed/emitted by the chain. The onsite potential term $\sum_n n \omega |n \rangle \langle n|$ tracks the energy of these quanta, but may be interpreted as an effective electric field. We discuss the eigenstates of the transformed model~\eqref{eq:frequency_lattice} and the Floquet states of the original model~\eqref{eq:model_eq} interchangeably, as they are related by a Fourier transformation. For a more detailed deviation of the synthetic lattice model see~\cite{PhysRevLett.126.106805_synth_lat}.

\begin{figure}
    \centering
    \includegraphics[width=\linewidth]{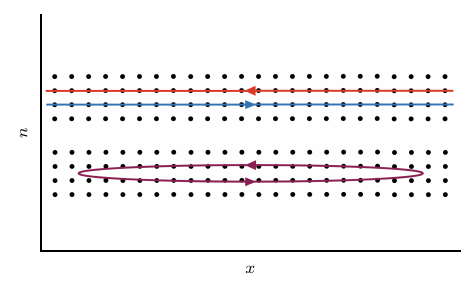}
    \caption{Sketch of quasienergy states on the synthetic lattice ($n$ number of quanta of the drive, $x$ labels sites of the chain) in the adiabatic limit (upper lattice) where the quasienergy states are delocalised and carry current in opposite directions. Any non-zero drive frequency causes hybridization of the quasienergy states (lower lattice).}
    \label{fig:qe_states_fig}
\end{figure}

The quantised pumping for $W<W_c$ and $\omega \to 0$ can be understood as a linear response to the weak electric field in the frequency lattice~\cite{PhysRevX.7.041008_synth_latt_qubit, PhysRevB.99.064306_synth_lat_qubit}.
Without the electric field and at weak disorder the system forms well defined bands of localised instantaneous states. In the presence of Berry curvature, a wavepacket prepared in the instantaneous lower band of the system without the electric field will respond by moving perpendicular to the electric field, which points along the $n$ axis. There are two families of states carrying currents in opposite directions along the spatial dimension, associated with the instantaneous lower and upper bands (Fig.~\ref{fig:qe_states_fig} upper lattice). Adding up the contributions to $\bar{Q}$ from all the states within each family, charge pumping is quantised. 

However, beyond linear response, the eigenstates in the synthetic lattice (Eq.~\eqref{eq:frequency_lattice}) are localised in both the spatial and synthetic dimensions. Thus, they do not carry current indefinitely. Indeed, the effective electric field localises eigenstates in the $n$ direction. This confines eigenstates to a 1D disordered slab, which must be localised by Anderson localisation~\cite{PhysRev.109.1492_anderson_loc, lee1985disordered}. More concretely, the counter-propagating states weakly couple to states with equal energy, causing hybridization (Fig.~\ref{fig:qe_states_fig} lower lattice). A state in the upper band can be resonant with one in the lower band that is shifted in the electric field direction by $d=\mathcal{O}\left( \Delta/\omega \right)$, with $\Delta$ denoting the onsite potential in the lattice. 
Due to Stark localisation, the matrix elements coupling the counter-propagating states are exponentially small in this distance---resulting in large localisation lengths at finite frequency, as above.

In the two-tone driven qubit, we replace the spatial dimension of the Thouless pump with another drive, incommensurately related to the first one. The electric field now points along a direction irrationally related to the lattice vectors, creating an inhomogeneous potential on sites closest to a given levelset~\cite{PhysRevB.105.144204_MBL_qp}. The linear response of the system is a quantised pumping of energy from one drive to the other. Away from linear response, the synthetic eigenstates also have $\mathcal{O} ( e^{1/\omega} )$ large localisation lengths. Localisation of the synthetic eigenstates implies that the average energy pumped between the drives is zero, but the manifestation of large localisation lengths are giant energy oscillations between them~\cite{PhysRevB.108.134303_qubit}.

\section{Comment on power law scaling in the non-adiabatic transition mechanism}

\begin{figure}
    \centering
    \includegraphics[width=\linewidth]{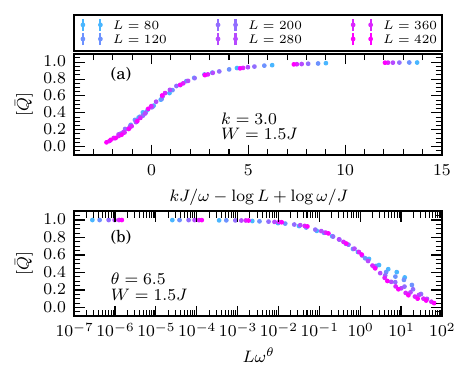}
    \caption{Scaling collapse of the disorder averaged $\bar{Q}$ with (a) $[\bar{Q}](\xi_{\omega},L) \sim [\bar{Q}](\xi_{\omega}/L)$ with $\xi_{\omega} \sim \frac{\omega}{J}e^{kJ/\omega}$ and (b) $[\bar{Q}](\xi_{\omega}',L) \sim [\bar{Q}](L/\xi_{\omega}')$ with $\xi_{\omega}' \sim \omega^{-\theta}$. \emph{Parameters}: $J=1.0$, $\Delta_0=1.5J$, $\delta_0=0.5J$, $W=1.5J$, (a) $k=3.0$, (b) $\theta=6.5$ and $[\bar{Q}]$ is averaged over 200 disorder realisations.}
    \label{fig:scaling_app_fig}
\end{figure}

Ref.~\cite{grabarits2023floquetanderson} also extracted the charge pumped in the steady state $\bar{Q}$ at finite system sizes, drive frequencies and disorder strengths. They conjectured that the scaling form for the total pumped charge is
\begin{equation}\label{eq:scaling_form_wrong}
    [\bar{Q}](\xi_{\omega}', L) \sim \Tilde{Q}(L/\xi_{\omega}'),
\end{equation}
where the length scale over which charge is pumped scales as a power law in drive frequency. $\xi_{\omega}' \sim \omega^{- \theta(W)}$. Although their conjecture contrasts our prediction, $\xi_{\omega} \sim e^{k J/\omega}$, the data reported in Ref.~\cite{grabarits2023floquetanderson} is consistent with our proposed scaling. 

The comparison between the scaling forms used in Fig.~\ref{fig:scaling_app_fig} (a) ($\xi_{\omega} \sim e^{kJ/\omega}$) and Fig.~\ref{fig:scaling_app_fig} (b) ($\xi_{\omega}' \sim L \omega^{\theta})$ show that the exponential form $\xi_{\omega}$ is the better fit to our data at small disorder strength $W = 1.5 J$. Moreover, the authors of Ref.~\cite{grabarits2023floquetanderson} were able to fit the exponential ansatz ($\xi_\omega$) to their data for $W<2J$ (see ~\cite[Appendix IC]{grabarits2023floquetanderson}). The large exponents ($\theta > 6$) extracted with power law scaling for $W<2J$ also indicate that an exponential fit performs better.

The authors of Ref.~\cite{grabarits2023floquetanderson} support a power law scaling form due to the behavior of \([\bar{Q}]\) at larger disorder strengths, \(W>2J\), which are closer to the critical point of \(W_c \approx 3.16 J\). Assuming an exponential scaling form, we note that the prefactor $k$ in $\xi_{\omega} \sim e^{k J/\omega}$ decreases as the disorder strength approaches its critical value, $W \to W_c$ (Appendix~\ref{sec:non-adiabatic}). Thus, smaller values of \(\omega\) are required to resolve the exponential scaling for \(W>2J\) than for \(W<2J\). We propose that the power law fits performed at larger disorder strengths in Ref.~\cite{grabarits2023floquetanderson} effectively probe the behaviour of the scattering rate~\eqref{eq:exc_rate_app} in the intermediate drive frequency regime. The scattering rate in Eq.~\eqref{eq:full_scattering_rate_appendix} only gives the exponential dependence in $1/\omega$ in the asymptotic limit of $\omega \to 0$.  Therefore, for the same range of $\omega$, the exponential behaviour is probed better at disorder strengths further away from the critical value. 
\end{document}